\documentclass[12pt]{article}
\usepackage{graphics,psfrag}
\usepackage{amsthm,amssymb,epsfig,amsmath,euscript,array,cite}

\newcommand{\be}{\begin{equation}}
\newcommand{\ee}{\end{equation}}
\def\bea{\begin{eqnarray}}
\def\eea{\end{eqnarray}}
\newcommand{\eq}[1]{(\ref{#1})}

\newcommand{\beq}{\begin{equation}}
\newcommand{\eeq}{\end{equation}}
\newcommand{\ben}{\begin{eqnarray}}
\newcommand{\een}{\end{eqnarray}}
\newcommand{\bes}{\begin{subequations}}
\newcommand{\ees}{\end{subequations}}
\newcommand{\blg}{\begin{align}}
\newcommand{\elg}{\end{align}}

\newcommand{\cN}{{\cal N}}

\def\one{\mbox{1 \kern-.59em {\rm l}}}

\def\a{\alpha}      
\def\b{\beta}

\def\e{\epsilon}

\def\k{\kappa}
 \def\L{\Lambda}
\def\m{\mu}

\def\s{\sigma}  
\def\t{\tau}
\def\th{\theta}

 \def\cH{{\cal H}} 
  
 \def\cN{{\cal N}}

\begin{document}

\hfill{WITS-CTP-062}

\vspace{20pt}

\begin{center}

{\Large \bf
Gluon Scattering Amplitudes in Finite Temperature Gauge/Gravity Dualities
}
\vspace{20pt}

{\bf
George Georgiou$^a$, Dimitrios Giataganas$^b$
}

{\em
${}^a$Demokritos National Research Center \\ Institute of Nuclear Physics\\
Ag. Paraskevi, GR-15310 Athens, Greece,\\
${}^b$ National Institute for Theoretical Physics,\\
School of Physics and Centre for Theoretical Physics,\\
University of the Witwatersrand,\\
Wits, 2050,\\
South Africa
}

{\small \sffamily
georgiou@inp.demokritos.gr,\,
dimitrios.giataganas@wits.ac.za
}

\vspace{30pt}
{\bf Abstract}
\end{center}
We examine the gluon scattering amplitude in $\cN=4$ super Yang-Mills at finite temperature with nonzero R-charge
densities, and in Non-Commutative gauge theory at finite temperature.
The gluon scattering amplitude is defined as a light-like Wilson loop which lives at the horizon of the T-dual black holes of the backgrounds we consider. We study in detail a special amplitude, which corresponds to forward scattering of a low energy gluon off a high energy one.
For this kinematic configuration in the considered backgrounds, we find the corresponding minimal surface which is directly related to the
gluon scattering amplitude. We find that for increasing the chemical potential or the non-commutative parameter, the on-shell action
corresponding to our Wilson loop in the T-dual space decreases. For all of our solutions the length of the short side of the Wilson loop
is constrained by an upper bound which depends on the temperature, the R-charge density and the non-commutative parameter. Due to this constraint,
in the limit of zeroth temperature our approach breaks down since the upper bound goes to zero,
while by keeping the temperature finite and letting  the chemical potential or the non-commutative parameter to approach to zero the limit is smooth.

\setcounter{page}0
\newpage

\section{Introduction}
Lately the gluon scattering amplitudes of $N=4$ SYM have attracted much attention.
This is due to two reasons. One reason is that they are part of the more complicated QCD amplitudes.
The second reason is that exhibit interesting symmetries and structures  \cite{Witten:2003nn,Cachazo:2004kj,Georgiou:2004wu,Georgiou:2004by,Drummond:2007aua}, 
such as the dual conformal invariance, which are not at all apparent from the Lagrangian of the system.
An important observation was that these amplitudes have a certain iterative structure at the two and three loop level
\cite{Anastasiou:2003kj,Bern:2005iz}. This allowed the authors of \cite{Bern:2005iz} to make a conjecture for the all-loop result of the
Maximally Helicity Violating (MHV) amplitudes with any number of external legs.
Subsequently, this conjecture was proved to be true only for amplitudes with four and five external particles.
Beyond that one has to modify the BDS ansatz by adding a function $R$ depending only on certain cross-ratios of the
kinematic invariants of the amplitude. As a result this remainder function $R$ exhibits a dual conformal invariance.
The deep reason behind this invariance is the intriguing relation between light-like Wilson loops\footnote{The Wilson loops in the AdS/CFT where firstly examined in \cite{maldacenawl,Reywl} and in the finite temperature in \cite{Brandhuberwl}. } and scattering
amplitudes of $N=4$ SYM \cite{AM}, first observed at strong coupling by using the AdS/CFT correspondence \cite{maldacena98,witten98}.
Shortly after that the same relation was verified to hold at weak coupling too \cite{Drummond:2007cf,Brandhuber:2007yx}.

Since then there is an increasing effort to understand better the gluon scattering amplitudes using the conjecture made by Alday and Maldacena.
One the directions followed is the study of scattering amplitudes in other theories than the $\cN=4$ super Yang-Mills.
Since the internal space bulk geometry as well as the fermionic sector of the theory do not appear to play any role
in the relation of the Wilson loops to the light-like Wilson loops in the strong coupling,
one expects that the results will not depend on deformations of the internal space.
For example in the $\beta$ deformed theories, with real deformation parameter,
it has been found that the scattering amplitudes are the same as in $\cN=4$ super Yang-Mills theory \cite{Khoze05,Oz07}.
This result is naturally expected also from the form of the Wilson loops in these deformed theories studied in \cite{giataganas07}.
The situation in similar for the orbifolds on $\cN=4$ super Yang-Mills where with a rescaling of the coupling constant the amplitudes remain
the same as the initial theory.
An other interesting generalization of the Wilson loop-gluon scattering amplitude conjecture is done in $\cN=2$ theory with matter
in the fundamental representation \cite{Komargodski07,Mcgreevy07} where it was shown that amplitudes for quarks and gluons
at strong coupling can be constructed from specific gluon amplitudes.
Moreover, an extension of the gluon scattering correspondence to the finite temperature seems to be a very important problem\cite{Nastasegs},
especially due to difficulties to define asymptotic states in the strong coupling scattering of the dual field theory.
It is expected that the results compared to the zero temperature case will be modified crucially, since the $AdS$ part of the metric
does change. Indeed, the T-dual background is not anymore
equal to the initial one, making the conjecture more complicated. Also due to the presence of a black hole, it is natural
to place the light-like Wilson loop at its horizon which in the T-dual background is located at the $UV$. As a consequence
of the complicated metric the minimal surface problem becomes highly non-trivial. Hence, until now only a special Wilson loop
configuration has been considered. This configuration has been interpreted as the forward scattering
of a low energy gluon off a high energy gluon. The claim is that this kinematic configuration corresponds to a light-like Wilson
loop which lives at the horizon of the T-dual back hole with one edge much longer than the another. Unfortunately, if one apply strictly
the extension of the Alday-Maldacena conjecture to the finite temperature this kinematic configuration gives only a phase contribution
to the amplitude as also noticed in \cite{Nastasegs}.

In this paper,  we continue and extend in many ways the study  of gluon scattering amplitudes at finite temperature.
We find this topic particularly interesting since apart from its high importance of solving it in full generality in the dual gravity
theory, a possible solution will give several hints for the corresponding  progress in the dual field theory.
In Section 2, we introduce the dual gravity backgrounds of the theories we will be using. We start with the gravity background dual to $N=4$ SYM at
finite temperature. Then we present the background dual to $N=4$ SYM at
finite temperature and finite chemical potential. Subsequently, we write down the gravity dual of non-commutative gauge theories at
finite temperature. In all cases we perform a series of four T-duality transformations along the four directions of the world-volume of the brane
on which the ends of the open strings which scatter live.

Having obtained the T-dual backgrounds we proceed to Section 3, where we calculate
the minimal surface for a Wilson loop which corresponds to the forward scattering of a low energy gluon ($E<T$) off a high energy one ($E>T$).
More precisely, in Section 3.1 we consider the case of gluon scattering amplitude at finite temperature,
in Section 3.2 the case of gluon scattering amplitude at finite temperature with finite chemical potential and in Section 3.3
the case of gluon scattering amplitude in non-commutative theories at finite temperature.
In all cases the amplitude turns out to be a pure phase times a prefactor which as in the original $AdS$ case should be the tree level amplitude.
This behavior is due to the Lorentzian nature of the world-sheet metric for this particular configuration.
We also find that as we increase the chemical potential the action corresponding to a specific Wilson loop living in the T-dual metric decreases.
Similarly, increase of the non-commutative parameter leads to a decrease of the on-shell action of the Wilson loop considered.
Furthermore, the energy of the low energy gluon should be lower than a critical value $L_{2max}$ for the solution to exist.
Put in another way, for any fixed value of the low energy gluon the temperature can not be smaller than a specific value.
This means that the zero temperature limit of our solutions does not exist.
On the contrary, setting to zero the chemical potential or the non-commutativity of the space is always a well-defined limit.

\section{The backgrounds}

Here we present the backgrounds which will be used in the following sections.
The common feature of all these backgrounds is that they are dual to strongly coupled field theories
at finite temperature. We start with the background that is dual to $\cN=4$ super
Yang-Mills at finite temperature. Then we consider the background dual to
$\cN=4$ super Yang-Mills at finite temperature and finite R-charge density
in order to capture some more dynamical properties of the gluon scattering amplitudes and their dependence on the temperature and chemical potential.
Finally, we  present the backgrounds dual to non-commutative gauge theories at finite temperature in order to investigate how the gluon scattering amplitudes depend on the non-commutative parameter.

The gluon scattering amplitude is defined as the scattering of open strings whose ends are located on an IR-brane sitting at the horizon of the corresponding geometry.
As in \cite{AM}, the boundary conditions are simplified by performing a T-duality on these backgrounds. After the T-duality the problem reduces to finding the expectation value of a light-like Wilson loop on the UV
of the T-dual geometry.
The T-duals of all the backgrounds we use here differ from the original ones, unlike the case of the $AdS_5$ where the T-dual background is also $AdS_5$. Moreover, in the non-commutative case the T-duality changes the
B-fields according to the T-duality rules. While the T-duality performed on the other two backgrounds does not generate any new B-fields.

The T-dual coordinates $y$ are related to the initial ones $x$ through an equation that has the following form
\ben\label{defcoord}
\partial_\a y^\mu=i w(r)^2 \e_{\a\b}\partial_{\b} x^\m
\een
where $w(r)$ is a function that can be found analytically and depends on the form of the metric.

Hence, for each of the following backgrounds we perform a 'disk' T-duality in the four coordinates of the $AdS$ part and specify the boundaries and the horizon of the new metrics.

$\bullet$ \textbf{Finite Temperature $\cN=4$  super Yang-Mills dual background.}

To introduce temperature in the AdS/CFT we need to deform the  $AdS_5$ background by placing a big black
hole in the center of the $AdS$ space \cite{Wittenft} or equivalently, to calculate the near horizon geometry of a
of near extremal black 3-brane. 
The resulting metric  is equal to
\ben\label{metricft}
ds^2= \frac{r^2}{R^2}(-h_T dx_0^2+dx_1^2 + dx_2^2 + dx_3^2)+\frac{R^2}{r^2 h_T} dr^2~,
\een
without any additional fluxes compared to the original $AdS_5$ background and where
\be\label{ht}
h_T=1-\frac{r_h^4}{r^4}~.
\ee
The boundary of this space is at $r=\infty$, while the horizon is situated in the
bulk at $r = r_h$. The 'disk' T-duality on this space does not generate
any new fields, although the new metric is not equal to the undeformed one.
After the T-duality the metric \eq{metricft} becomes
\ben\label{Tmetricft}
ds^2=\frac{R^2}{r^2}\left(-\frac{dy_0^2}{h_T}+dy_1^2 + dy_2^2 + dy_3^2+\frac{dr^2}{h_T}\right)
\een
where $y$ are the T-dual coordinates. The T-dual metric  has its horizon at $r=r_h$ and boundary located at $r=0$. We will place our
light-like Wilson loop at $r=r_h$, which makes sense to do so since the point $r=r_h$ is at the UV of the T-dual metric as also commented in \cite{Nastasegs}. 

$\bullet$ \textbf{Finite Temperature $\cN=4$  super Yang-Mills dual background with chemical potential.}

To introduce a chemical potential in the previous background we need to introduce non-zero R-charges, which corresponds to rotating near extremal D3-branes.
These backgrounds are characterized by  a non extremality parameter and in general by three rotation parameters which can be though as corresponding to chemical potentials.
The background we consider here has two non zero angular momentum parameters which we set equal to each other \cite{Sfetsosch2} and is a special case of \cite{Sfetsosch1}. The metric reads
\ben\nonumber
&&ds^2= H_{ch}^{-1/2} \left[-\left(1-\frac{r_h^4 H_{ch}}{R^4}\right) dx_0^2 + dx_1^2 + dx_2^2 + dx_3^2\right]
+ H_{ch}^{1/2} \frac{r^4(r^2-\m^2 \cos^2\th)}{(r^4-r_h^4)(r^2-\m^2)} dr^2\\ \nonumber && \hspace{2.5cm}+H_{ch}^{1/2}
\Big[(r^2-\m^2\cos^2\th)d\th^2+r^2\cos^2\th d\Omega_3^2+\left(r^2-\m^2\right)\sin^2\th d\phi_1^2\\
&& \hspace{5.7cm}~~-2\frac{r_h^2 \m}{R^2}\cos^2\th ~dt\left(\sin^2\psi d\phi_2+\cos^2\psi d\phi_3\right)\Big]~,\label{ftcpmetric}
\een
where
\ben
H_{ch}=\frac{R^4}{ r^2 (r^2-\m^2\cos^2 \th)}~.
\een
The horizon of the geometry \eqref{ftcpmetric} is located at $r=r_h$
while the boundary of the space is at $r=\infty$.
The parameter $r_h$ of the metric is a non-extremality parameter, and $\m$ is the common value of two of the three rotation parameters which
correspond to the three generators of the Cartan subalgebra of $SO(6)$. The third one we set it equal to zero.
This particular choice has nothing special and is enough to capture all the essential features of gluon scattering amplitudes in finite temperature and R-charge density.

We will choose to work in the Grand Canonical Ensemble (GCE) where the thermodynamic quantities which
we keep constant are the temperature T and the chemical potential $\hat{\m}$. These are related to the
non-extremality and rotation parameters by
\ben
T=\frac{\sqrt{r_h^2-\m^2}}{ \pi R^2}~,\qquad\mbox{and}\qquad \hat{\m}=\frac{\m}{R^2},
\een
where $r_h\geq \m$, to require real temperature. However, the stability conditions of the black hole require the stricter inequality
\be\label{constr1}
r_h\geq\sqrt{2}\m~. 
\ee
By performing the necessary for our purpose disk T-duality on this metric we get
\ben\nonumber
ds^2&&= H_{ch}^{1/2}\left(-\frac{1}{1-\frac{r_h^4 H_{ch}}{R^4}}dy_0^2+dy_1^2 + dy_2^2 + dy_3^2\right)+H_{ch}^{1/2}\frac{r^4\left(r^2-\m^2\cos^2\th\right)}{
\left(r^4-r_h^4\right)\left(r^2-\m^2\right)}dr^2\\&&\hspace{7.8cm}+H_{ch}^{1/2}
\left[(r^2-\m^2\cos^2\th)d\th^2\right]~,\label{tftcptmetric}
\een
where 
 the horizon remains at $r=r_h$, and in the above metric we present only the metric elements relevant to our study.

$\bullet$ \textbf{Background dual to noncommutative
Yang-Mills theory at finite temperature.}

To construct this background we introduce the B-field to the non-extremal D3 brane background using U-duality \cite{Maldacenanc,Hashimoto99,Jabbari99}. The result
reads
\ben\label{metricnc}
ds^2&=&\a'R^2\left(r^2\left(-h\left(f dx_0^2+dx_1^2\right)+h\left(dx_2^2+dx_3^2\right)\right)+\frac{dr^2}{f r^2}+d\Omega_5^2\right)~,\\
B_{01}&=&\a'R^2 a^2 r^4 h~,\qquad B_{23}= \a'R^2 a^2 r^4 h~,
\een
where
\ben
f=1-\frac{r_h^4}{r^4}~,\qquad\mbox{and}\qquad h=\frac{1}{1+a^4 r^4}~.
\een
$a$ is the non-commutative parameter and the horizon is located at $r=r_h$. It is evident from \eqref{metricnc} that
the  non-commutativity in dual field theory has introduced only between the $(x_0,x_1)$
and $(x_2,x_3)$. A second remark concerns the signature of the metric above which is chosen to be $(-2,2)$ and we need to take that in account for the ansatz of the world-sheet we consider.
To calculate the gluon scattering amplitude  we need as usual to perform a T-duality on the four $x_i$ coordinates and find in
the new background the appropriate minimal surface.

After performing the T-dualities  the new background becomes
\ben\label{nctmetric}
&&ds^2=a' R^2\left(\frac{-1}{h^2 r^2\left(f+a^4 r^4\right)}\left(dy_0^2+f dy_1^2\right)+\frac{1}{r^2}\left(dy_2^2+dy_3^2\right)+\frac{dr^2}{r^2 f}+d\Omega_5^2~\right),~~~~~\\\label{nctb}
&&B_{01}=-a' R^2 \frac{a^2}{h(f+a^4 r^4)}~,\qquad B_{23}=- a' R^2 a^2 ~,
\een
As in the previous case, 
the horizon of the T-dual metric is situated at $r=r_h$, and it is
where we place the Wilson loop whose expectation value will give the gluon amplitude at strong coupling.

\section{Gluon scattering amplitudes at finite temperature}



The aim of this section is to examine the behavior of gluon scattering amplitudes in the presence of a heat bath of temperature T.
The scattering of the gluons at finite temperature is described by the scattering of open strings whose ends are attached to the a IR-brane of the corresponding geometry, eg. \eqref{metricft}, as in the original Alday-Maldacena construction \cite{AM}.
However, here the existence of the horizon puts a limit to the position of the brane. As also done in \cite{Nastasegs}, we choose to place the IR-brane exactly on the horizon. After performing the T-dualities , as described in the previous section, we end up with a background \eq{Tmetricft} which describes the same physics with the initial one.
The advantage is that the boundary conditions of the problem simplify significantly. Namely, the whole problem reduces to the calculation of a Wilson loop expectation value, which lives at the horizon of the black hole in the T-dual background.
This Wilson loop consists of light-like segments whose length will be proportional to the momenta of the scattered gluons.

Two comments are in order. Firstly, it may seem unusual
that we consider the scattered strings on a brane sitting at the IR of the original geometry.
Usually, observables in the context of AdS/CFT are defined on the boundary of the space, i.e. at $r=\infty$.
The resolution of this puzzle is that the IR-brane does touch the boundary of the space at $t=\pm \infty$ \cite{Nastasegs}. Consequently, the endpoints of the asymptotic string states are on the boundary of the gravity dual. Secondly, it does makes sense to place a Wilson loop at the black hole horizon since in the T-dual metric the horizon is sitting at the UV region of the dual geometry.

Unfortunately it is very difficult to find the minimal surface  with the most general boundary conditions even for the case of four gluons. This is due to the complicated T-dual metric.
Furthermore, through  the introduction of temperature we have also lost integrability whose role was instrumental in obtaining the area of the minimal surface in the case of pure $AdS$. Thus we will limit ourselves to a Wilson loop with particular boundary conditions.
Namely, we consider a rectangular loop with one edge much bigger than the other. We call the long edge of this Wilson loop  $L$ and the short one $L_{2}$, where $L\gg L_{2}$, with both of them being in light-like direction. 
This light-like Wilson loop corresponds to an amplitude at strong coupling for forward scattering of a 'low' energy gluon off a high energy gluon.

Since one edge of the loop is much larger than the other we make the following ansatz for the embedding of the world-sheet
\footnote{Our ansatz differs from the one of \cite{Nastasegs}. Here we have allowed $y_0$ to depend on $\sigma$, in order to satisfy our equations of motion for $y_0$.}
\be\label{ansatz}
y_0=\tau +f(\s)~,\qquad y_1=\t~,\qquad y_2=\sigma~,\qquad r=r(\sigma),
\ee
with the suitable boundary conditions
\ben\nonumber
&&y_0=\tau~,\hspace{2.cm} y_1=\tau~,\qquad y_2=0\qquad \mbox{at}\hspace{1.2cm} \s=0~,\\
&&y_0=\tau+\frac{L_2}{\sqrt{2}}~,\qquad y_1=\tau~,\qquad y_2=\frac{L_2}{\sqrt{2}}\qquad \mbox{at}\qquad \s=\frac{L_2}{\sqrt{2}}~.\label{ybc}
\een
They impose that the worldsheet should end on the two long light-like edges of the Wilson loop. Similarly, the requirement that the worldsheet should also end on the two short light-like edges of the Wilson loop gives
\ben\nonumber
&&y_0=\sigma~,\hspace{2cm} y_1=0~,\hspace{1.2cm} y_2=\sigma \qquad \mbox{at}\qquad \tau=0~,\\
&&y_0=\sigma+\frac{L}{\sqrt{2}}~,\qquad y_1=\frac{L}{\sqrt{2}}~,\qquad y_2=\sigma \qquad \mbox{at}\qquad \tau=\frac{L}{\sqrt{2}}~.\label{ybcshort}
\een
The factor of $1/\sqrt{2}$ comes from the projected coordinate of the relevant edge of the Wilson loop to the $y_0$ axis.
In what follows, we will relax the second set of boundary conditions \eqref{ybcshort}.
The rationale behind this is the following. Since one edge is much larger than the other $L>>L_2$ one
can assume that the world sheet is translationally invariant in the $L$ direction, i.e. it does not depend on $\tau$.
If this is the case then it is obvious that \eqref{ybcshort} can not be satisfied.
What really happens is that the worldsheet remains independent of $\tau$ almost everywhere except very close to
the short edges where it rapidly falls to end on the lines defined by \eqref{ybcshort}.
The contribution to the total area of this small region near the short edges is negligible due to
the condition $L>>L_2$. This justifies the ansatz of \eqref{ansatz}.

One important comment is in order. One may argue that the solution described in the last paragraph does not exist
or if it exists it is not the one with the minimal action. One could imagine, for example, a solution which
oscillates many times up and down between the two short edges\footnote{However, our Wilson loops here are defined at finite $r$, and the situation is different than the \cite{Argyresjet,Liujet}.}. Such a solution, if it exists, clearly does not satisfy
\eqref{ansatz}. We have not been able to exclude such solutions and we will proceed assuming that a solution of the type \eqref{ansatz} exists and that it is the dominant one, since we do not have opposite indications.

 Before we proceed in the next section it is convenient to introduce some notation. The factor of the amplitude that can be calculated from the AdS/CFT is of the form $A\sim \exp(i S)$ where $S=i S_{E}\sim\sqrt{-\det g}$ and $S_E\sim\sqrt{\det g}$. The kinematic configurations we consider for the three different theories below turn out to have Lorentzian signature world-sheet, and hence the results on the amplitude are of the form $A\sim \exp(i S)$, where $S$ is the real  on-shell action which depends on the parameters of the  theory and the configuration.

In the following subsections
we calculate the gluon scattering amplitudes in three theories with different characteristics, using the set up we presented here.

\subsection{Gluon scattering amplitude in finite temperature \cN=4 supersymmetric Yang-Mills}

~~~~In  this section we consider the gluon scattering amplitude in the gravity dual background of the finite temperature $\cN=4$ supersymmetric Yang-Mills.
For the world-sheet we use the configuration \eq{ansatz} and the relevant boundary conditions \eq{ybc}.


The Nambu-Goto action in the static gauge for our world-sheet becomes
\be\label{NGT}
S=\frac{1}{2 \pi\a'}\int dy_1 dy_2\frac{R^2}{r^2}\sqrt{-\left(1+\frac{(\partial_i r)^2-(\partial_i y_0)^2}{h}-\frac{\partial_0 r \partial_1 y_0-\partial_1 r \partial_0 y_0}{h^2}\right)}~,
\ee
where for this section $h=h_T$ given by the equation \eq{ht}.
The equation of motion for $y_0$ gives an equation for the $f'$:
\be\label{fprime}
f'=\frac{c_1 h r^2 \sqrt{D}}{R^2}~,
\ee
where
\be
D:= -\left(1+\frac{(\partial_i r)^2-(\partial_i y_0)^2}{h}-\frac{\partial_0 r \partial_1 y_0-\partial_1 r \partial_0 y_0}{h^2}\right)~,
\ee
and $c_1$ is the integration constant.
The Lagrangian density does not depend explicitly on $y_2$, which can be considered as the ''time'' in our system, so the Hamiltonian of this motion is constant. We set for convenience the Hamiltonian equal to $1/\b$,  so the factor $\b$ to appear in the numerator in the equations of turning points, as we will see.

Using the conjugate momenta
\be
p_{y_0}=\frac{R^2 f'}{r^2 h \sqrt{D}}~,\qquad p_r=\frac{R^2 r'}{r^2 h \sqrt{D}}\left(1-\frac{1}{h}\right)~,
\ee
the Hamiltonian reads
\be\label{Ham1}
\cH=\frac{R^2}{r^2 \sqrt{D}}\left(1-\frac{1}{h}\right)=:\frac{1}{\beta}~.
\ee
By eliminating $\sqrt{D}$ between equation \eq{fprime} and equation \eq{Ham1}, the $f'$ equation takes the simple form
\be\label{fprime2}
f'=-\b c_1 \frac{r_h^4}{r^4}~.
\ee
We can substitute $f'$ in \eq{fprime} to solve for $r'$, which gives
\be\label{rprime}
r'^2=\frac{1}{r^8}(-r^8+ b r_h^4 r^4 + c r_h^8)~,
\ee
with
\be
b:= 1- \b^2 c_1^2~,\qquad c:= \b^2(\frac{R^4}{r_h^4}+\frac{c_1^2}{2})~.
\ee
Now the turning points of the worldsheet can be found from the solutions of the equation \eq{rprime} to be
\be\label{turning}
r_\pm=\frac{r_h^4}{4}\left(b\pm \sqrt{b^2+4 c}\right)~.
\ee
Since we place the Wilson loop  on the horizon ($r=r_h$) the worldsheet we consider extends from $r=r_h$ to $r=r_+$. The solution $r_-$ is negative and as such is physically irrelevant.

The equation \eq{fprime2}  and  \eq{rprime}  can be rewritten as
\ben\label{fprime3}
\int_{r_h}^r dr \frac{-\beta c_1 r_h^4}{\sqrt{-r^8+ b r_h^4 r^4 + c r_h^8}}=\int_{0}^f  d f~
\een and
\ben\label{rprime2}
\int_{r_h}^r dr \frac{r^4}{\sqrt{-r^8+ b r_h^4 r^4 + c r_h^8}}=\int_{0}^{\sigma} d\sigma~.
\een
and can be integrated analytically in terms of Appell hypergeometric functions $F1$ of two variables, which we call AppellF1.
The final system we find has two unknown integration constants: $\b$ and $c_1$. To specify them we need to impose the
boundary
conditions given above. Then the integration of \eq{fprime3} and \eq{rprime2} respectively gives
\ben\label{lr}
\frac{L_2 \sqrt{2}}{4}&=&\left(\frac{r\sqrt{1-K+\L}}{\sqrt{K\L}} \mbox{AppellF1}\left[\frac{5}{4},\frac{1}{2},\frac{1}{2},\frac{9}{4},
\frac{r^4}{r_+^4},\frac{r^4}{r_-^4}\right]\right)\bigg|_{r_h}^{r_+}~,\\\label{lf}
\frac{L_2 \sqrt{2}}{4 }&=&\left(\frac{r^5}{5\sqrt{K \L}r_h^4}\mbox{AppellF1}\left[\frac{1}{4},\frac{1}{2},\frac{1}{2},\frac{5}{4},
\frac{r^4}{r_+^4},\frac{r^4}{r_-^4}\right]\right)\bigg|_{r_h}^{r_+}~,
\een
where $K:=r_{+}^4/r_h^4$ and $\L:=-r_{-}^4/r_h^4$.
Notice that we have managed to write in the above equations everything in the RHS  in terms of $K$ and $\L$ since $b$ and $c$ and thus $\beta$ and $c_1$ can be expressed in terms of $K$ and $\L$ through \eq{turning}.
Hence, in principle one can solve  the system of \eq{lr} and \eq{lf} for $K$ and $\L$ and subsequently for $\beta$ and $c_1$. This can be done only arithmetically by giving values to $L_2$ and finding the corresponding values of $\b$ and $c_1$.

By solving the system we obtain the solutions in form of triplets $(L_{2},\,\b,\,c_1)$ and from \eq{lr} we get the dependence of $L_{2}$ on $r_+$. The result is plotted in Figure 1.
\begin{figure}[t]
\centerline{\includegraphics[width=75mm]{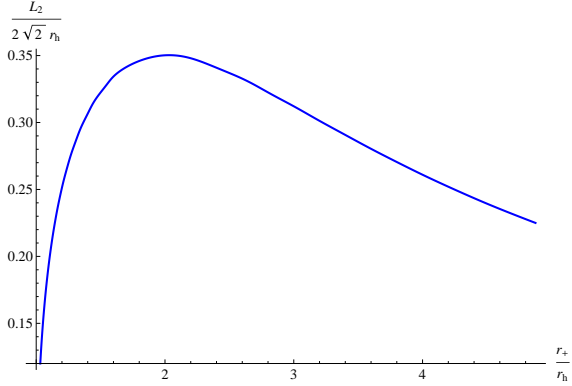}}
\caption{{\small The relation between $L_{2}/r_h$ and the $r_+/r_h$. There exist two solutions for the same  $L_{2}$, but one of them is acceptable.}}
\end{figure}
We find that for $L_{2\,max}:=L_{2}/(2 \sqrt{2} r_h)\simeq 0.35$ the curve has a maximum value at $K_{max}^{1/4}:=r_{+Lmax}/r_h\simeq 1.973$. Moreover, for each value of $L_{2}<L_{2\,max}$ there exist two different solutions for $r_+$. For the solutions corresponding to $r_+>r_{+Lmax}$ the string world-sheet goes deeper in the bulk compared to the corresponding ones with the same $L_{2}$ value where $r_+<r_{+Lmax}$.
Only one of these two sets of solutions should be acceptable and correspond to the gluon scattering amplitude. It is natural to think that is the one with the minimal action, but to decide we need to regularize the action and observe the behavior of each of these branches as $L_{2}\rightarrow 0$.

The corresponding minimal surface is calculated analytically as
\be\label{actionf}
S=\frac{R^2 L}{\sqrt{2} \pi \a' r_h}\int_{1+\epsilon}^{K^{\frac{1}{4}}} dz\frac{1}{z^4-1}\sqrt{\frac{c+ b-1}{c+z^4(b-z^4)}}~.
\ee
It is immediate to notice that this action is logarithmically divergent. This is
consistent with the fact that the gluon scattering amplitudes are IR divergent even at
finite temperature.  
To isolate the divergence one should regulate the action by performing the integral from $z=1+\epsilon$, since the divergence comes from the lower limit of the integral in \eq{actionf}.
The divergent part reads
\ben\label{div}
S_{div}=\frac{R^2 L}{\sqrt{2} \pi \a' r_h}\int_{1+\epsilon}dz \frac{1}{4(z-1)}=
-\frac{R^2 L}{4 \sqrt{2} \pi \a' r_h}\log{\epsilon}~.
\een
Then the finite part is given by
\be\label{regfin}
S_{fin}=S-S_{div}~
\ee
and for the solutions we found we can integrate the Lagrangian and get the dependence of the on-shell action from $r_+$. This is what we draw in the first plot of Figure 2, where we consider $\epsilon=10^{-5}$.
The branch which is physically acceptable is the one that satisfies the physical condition that the action $S(L_{2})$ should tend to zero as $L_{2}\rightarrow 0$ \cite{Nastasegs}.
By inspecting Figure 1 and the right plot of Figure 2 we see that this happens for $r_+/r_h\rightarrow 1$ which means that the branch
with $r_+<r_{+Lmax}$ is the physical one.
As expected, it turns out that for this set of solutions $r<r_{+Lmax}$, gives minimal action for any value of $L_{2}$ compared to the corresponding solutions for $r_+>r_{+Lmax}$ (Figure 2).
\begin{figure}[t]
\centerline{\includegraphics[width=70mm]{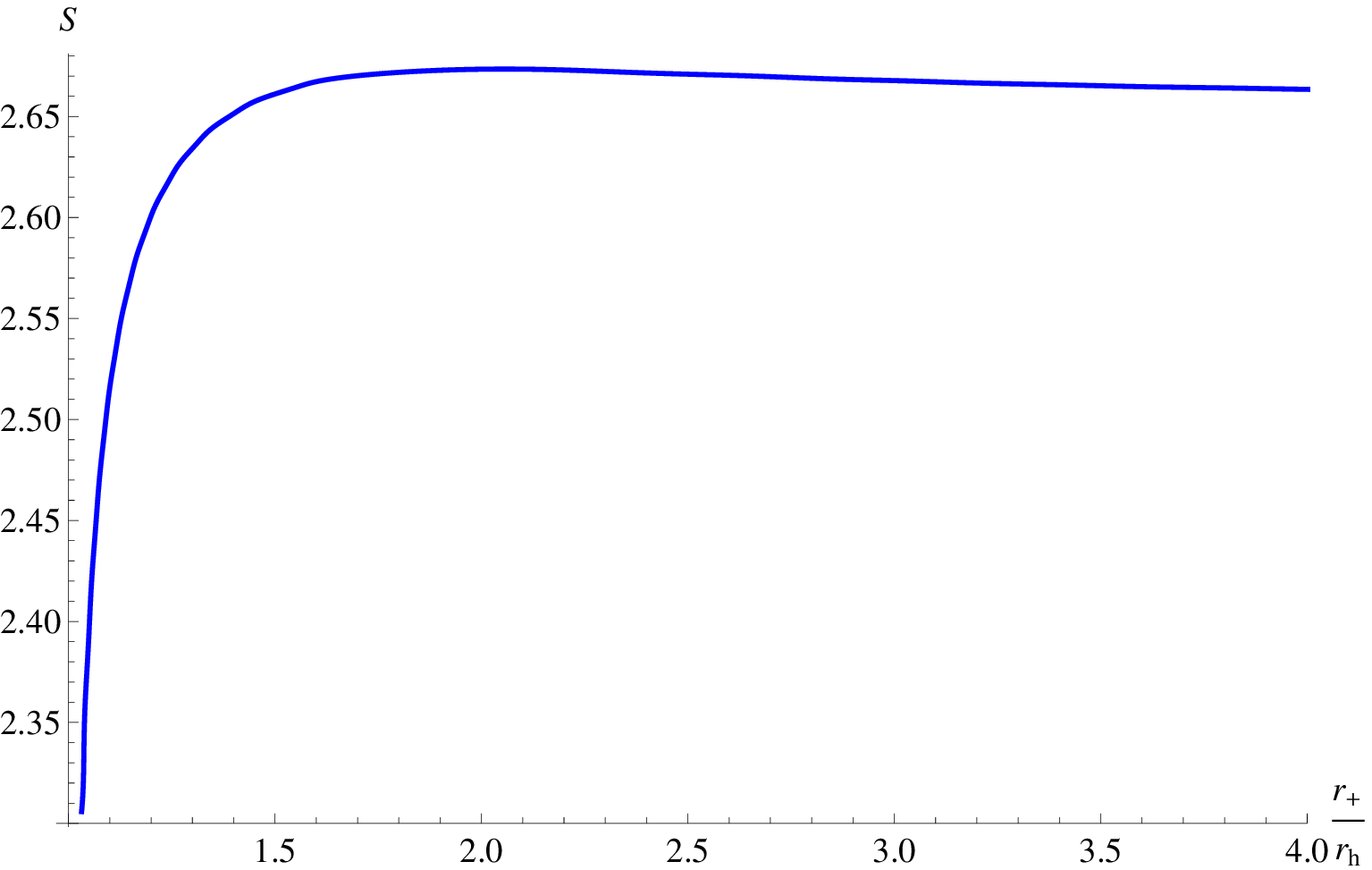}\quad\includegraphics[width=75mm]{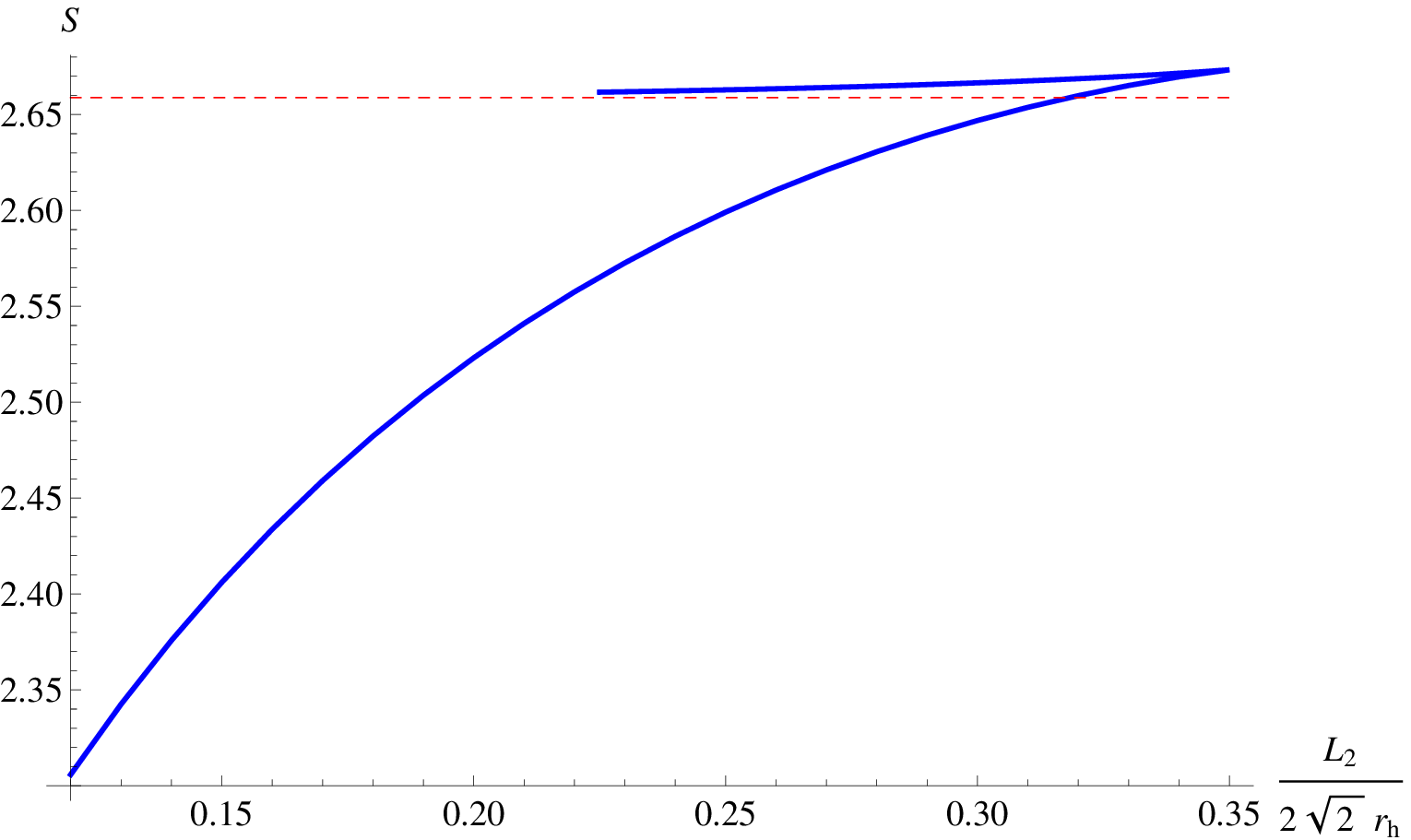}}
\caption{{\small On the left, the dependence of the action as a function of $r_+/r_h$. The action is normalized, such that the $S$ we draw is actually the action divided by $2 \pi \a'/(\sqrt{2} R^2 L)$. For $r_+<r_{+Lmax}$ the action is increasing. When it takes its maximum value it starts a slow decrease and tends to take asymptotically a constant value.  On the right, the dependence of the action as a function of $L_{2}$. The physical solution is the one that gives minimal action and corresponds to values of $r_+<r_{+L\,max}$. The red-dashed line correspond to the non-acceptable solution of the two parallel disconnected sheets.}}
\end{figure}

We close this section with a remark. There is another solution
to the equations of motion derived from the action \eq{NGT} with the same boundary conditions \eq{ybc}. This solution describes two parallel disconnected sheets which extend from $r=r_h$ to $r=\infty$. For this solution $f'=0$ and the corresponding action is
\ben\label{2lines}
S_d=\frac{1}{ \pi a'}\int d\sigma d\tau\frac{R^2 r_h^2}{r^4(1-\frac{r_0^4}{r^4})}r'&=&
\frac{\sqrt{\lambda}}{\pi}\int_{r_h(1+\epsilon)}^{\infty}dr\int d\tau\frac{r_h^2}{r^4-r_h^4}\nonumber \\
&=&-\frac{\sqrt{\lambda}}{\pi}\frac{L}{4\sqrt{2}r_h}(\log{\frac{\epsilon}{2}}+\frac{\pi}{2}) .
\een
As above the action is logarithmically divergent and one should take the lower limit of integration to be $r_h(1+\epsilon)$ and not $r=r_h$ in order to regulate the divergence. Subtracting the divergence from the action
we get its finite part
\ben\label{2linesf}
S_{dfin}=\frac{\sqrt{\lambda}}{\pi}\frac{L}{4\sqrt{2}r_h}(\log{2}-\frac{\pi}{2})
\een
and is the horizontal line in the second plot in Figure 2, which actually crosses the vertical axis and that the upper line of the $S(L_2)$ at the same point.
It is immediate to notice that the regularized action \eq{2linesf} depends only  on the long
edge of the Wilson loop $L$ and not on the short one $L_{2}$. 
We should note that this solution exists for any value of $L_{2}$, while the solution corresponding to the connected world-sheet exists only for $L_{2}<L_{2max}$. For very small values of $L_{2}$ the latter has smaller action and dominates the disconnected one. However, there is a critical point $L_{2crit}\simeq 0.3185$ beyond which the disconnected solution is the dominant one. This is similar to a first order phase transition of the Gross-Ooguri type. For small values of $L_{2}$ the favorable configuration is the one where the world-sheet penetrates into the bulk up to a finite point. Beyond the critical value $L_{2crit}$ the favorable configuration is the one where the world-sheet goes all the way to infinity where one can imagine that it merges with itself.

However this disconnected solution does not seem to have the physical requirements to be interpreted as a gluon scattering amplitude. On the other hand, it gives as a hint that could exist other kinds of solutions with similar boundary conditions that become dominant in different energies. Or even solutions that are valid for bigger values of $L_{2max}$. This could be related to the conjectured existence of different saddle points that become dominant at high energies when the string theory is expanded around them \cite{Grossmende}. Hence one could observe similar  phase transitions in the gluon scattering amplitudes as a reflection of the change in the character of the amplitudes due to different saddle point become dominant.

\subsection{Gluon scattering amplitude in finite temperature $\cN=4$ supersymmetric Yang-Mills with chemical potential}

In this section we study the relation of the gluon scattering amplitude on the chemical potentials at different temperatures. The background we use is \eq{tftcptmetric} and it comes by doing the disk T-dualities to the background \eq{ftcpmetric}.
The light-like Wilson loop and the ansatz for its worldsheet is similar to the finite temperature case \eq{ansatz},
with the same boundary conditions \eq{ybc}.

The Nambu-Goto action for this configuration becomes
\be\label{schem}
S=\frac{R^2}{2\pi\a'}\int dy_1 dy_2 \sqrt{H}\sqrt{-1-G_{00}\left(1+f'{}^2\right)-G_{rr}\left(G_{00}+1\right)r'{}^2}
\ee
where for convenience we define
\ben
H=H_{ch}\,,\qquad G_{00}=-\left(1-\frac{r_h^4 H}{R^4}\right)^{-1}\,,\qquad G_{rr}=\frac{r^4\left(r^2-\m^2\cos^2\th\right)}{
\left(r^4-r_h^4\right)\left(r^2-\m^2\right)}~.
\een
To satisfy the equation of motion for $\theta$ we choose
\ben
\th=\frac{\pi}{2}~.
\een
Then the equation of motion for $f'$ reads
\ben\label{fchprime}
f'=\frac{\sqrt{D} c}{H^{1/2}G_{00
}}~,
\een
where we have defined
\ben
D:=-\left(1+G_{00}\left(1+f'{}^2\right)+G_{rr}\left(G_{00}+1\right)r'{}^2\right)~.
\een
According to our boundary conditions \eq{ybc}, we observe from the equation \eq{fchprime} that the constant $c$ should be negative, since $G_{00}<0$.
We continue our analysis using the Hamiltonian formalism. The conjugate momenta read
\be
p_f=-\frac{G_{00}  f'\sqrt{H}}{\sqrt{D}}\,,\qquad\mbox{and}\qquad p_r=-\frac{G_{rr}\left(G_{00} +1\right)r'\sqrt{H}}{\sqrt{D}}~,
\ee
and the Hamiltonian then is equal to
\be\label{hcha}
\cH= \frac{\sqrt{H}\left(1+G_{00}\right)}{\sqrt{D}}~,
\ee
 which does not depend on $y_2$. Hence we can set it equal to a constant $1/\b$ and solve for $D$, which gives
\be\label{dchp}
\sqrt{D}= \sqrt{H} \b (1+G_{00})=\frac{\b r_h^4 R^2}{r_h^4 r^2-r^6}~.
\ee
Notice from \eq{hcha} that the constant $\b$ takes negative values since the $1+G_{00}<0$.
So the equation \eq{fchprime} can be  simplified to
\be\label{fchprime2}
f'=\frac{c \,\b\left(1+G_{00}\right)}{G_{00}}= \frac{\b c r_h^4}{r^4}~
\ee
and will be used to determine $\b$ and $c$.\footnote{The difference appearing in the sign between \eq{fprime2} and \eq{fchprime2}, is because the integration constant $c_1$ of equation \eq{fprime2} is positive so there $-\b c_1>0$, while here the integration constant $c$ is negative, so $\b c>0.$}
The turning point equation can be found by taking the square of the \eq{dchp} and using \eq{fchprime2}. The result reads
\ben\nonumber
r'^2&=&\left[-H \b^2 \left(1+G_{00}\right)-\left(1+\frac{c^2 \b^2\left(1+G_{00}\right)}{G_{00}}\right)\right]\frac{1}{G_{rr}}
\\\label{rprimech}&=&
\frac{\left(\left(r_h^4-r^4\right) \left(\b^2 c^2 r_h^4+r^4\right)
+\b^2 r_h^4 R^4\right) (r^2-\m^2) }{r^{10}}~,
\een
and by setting it equal to zero we find that the turning points are
\ben
&&r=r_h\,,\qquad\mbox{or}\qquad r= \m\,,\qquad \mbox{or}\\
&&r_+=r_h\left(1 + \frac{1}{2} \b^2 \left(-\left(\frac{1}{\b^2} + c^2\right) +\sqrt{\left(\frac{1}{\b^2} + c^2\right)^2 + \frac{4  R^4}{\b^2 r_h^4}}\right)\right)^{1/4}.\label{ttch}
\een
with the last solution to be acceptable, since the horizon is located at $r=r_h$. We observe that the turning point depends on the parameter $r_h$ monotonically, and as we will see the world-sheet enters deeper in the bulk and is more extended for increasing $r_h$.

The next step is to find the values of the parameters $\b$ and $c$ for each value of $L_{2}$ by solving the system of the following equations which come from integration of \eq{fchprime2} and \eq{rprimech} respectively
\ben\label{ch1}
&&\frac{L_{2}}{2\sqrt{2}}=\int_{r_h}^{r_{+}} \frac{c~ \b(1+G_{00})}{G_{00} ~r'}dr=\int_{1}^{\frac{r_{+}}{r_h}} \frac{\b c r_h r ~dr}{\sqrt{\left(\left(1-r^4\right) \left(\b^2 c^2 +r^4\right)+\frac{\b^2 R^4}{r_h^4}\right) (r^2-\k^2)}},~~~~\\\label{ch2}
&&\frac{L_{2}}{2\sqrt{2}}=\int_{r_h}^{r_{+}}\frac{1}{r'}dr=\int_{1}^{\frac{r_{+}}{r_h}} \frac{ r_h r^5~ dr}{\sqrt{\left(\left(1-r^4\right) \left(\b^2 c^2 +r^4\right)+\frac{\b^2 R^4}{r_h^4}\right) (r^2-\k^2)}}
\een
where $r'$ is given by \eq{rprimech} and we have made the change of variables $r\rightarrow r/r_h$. Instead of the $r_h$ we can insert in the above expressions the temperature and a dimensionless constant as
\be
r_h= T R^2 \sqrt{\pi^2+\xi^2}~.
\ee
The variables $\xi$ and $\k$  are defined as
\be
\xi:=\frac{\hat{\m}}{T}~, \quad \mbox{and} \quad \k:=\frac{ \hat{\m} R^2}{r_h}= \sqrt{\frac{\xi^2}{\xi^2+ \pi^2 R^2}}~.
\ee
Hence eventually the two integrals giving the $L_2$ can be written in terms of temperature and the dimensionless parameter $\xi$.
Then by solving appropriately the system of equations \eq{ch1} and \eq{ch2}, we can obtain a triplet of values $(\b,\,c,\,L_{2})$, for each pair of values of $T$ and $\xi$. We will keep fixed the temperature and vary the dimensionless parameter $\xi$ in order to examine the behavior of the gluon scattering amplitude in presence of chemical potential.

We also need to present  the analytic expression of the action which we also use in the numerical analysis. It can be written as
\ben\label{chs}
S=\frac{L}{\sqrt{2}\pi\a'}\int_{1}^{\frac{r_{+}}{r_h}} dr
\frac{-\b R^4 r}{r_h^3(r^4-1)\sqrt{\left(\left(1-r^4\right) \left(\b^2 c^2 +r^4\right)+\frac{\b^2  R^4}{r_h^4}\right) (r^2-\k^2)} }~,
\een
where we have used \eq{fchprime2} and \eq{rprimech}.
Unfortunately  the integrals \eq{ch1}, \eq{ch2} and \eq{chs} can not be done analytically, unlike the case with zero R-charge.

Notice that we could redefine the integration constants as $\b_1=\b R^2/r_h^2$ and $c_1=c r_h^2/R^2$, which would modify the turning point \eq{ttch} and the integrals \eq{ch1}, \eq{ch2} and \eq{chs}, to have only an overall $T$ dependence where the rest of the integral would depend only on $\xi$. Hence for fixed temperature we can solve the system \eq{ch1}, \eq{ch2} in terms of $L_2/T$ to obtain the constants $\b_1,\,c_1$. Then the action \eq{chs} would have an overall $R^4 L/(\a' T^3)$ dependence. So to answer the question of how the gluon scattering amplitude behaves in the presence of chemical potentials, is natural to compare the gluon scattering amplitudes  by keeping the temperature fixed and vary the parameter $\xi$.
Different values of the parameter $\xi$  result different values of the turning point of the world-sheet, and different solutions  $(\b,\,c,\,L_{2})$ since the system of equations \eq{ch1} and \eq{ch2}, depend on the dimensionless parameter $\xi$. Notice that it turns out that for some of the solutions $(\b,c)$, the turning point $r_+$ depends on $r_h$ almost in a linear way for different values of $\xi$.

For each case we examine, we will draw the plots $(L_{2}/T,\,r_+/r_h)$, $(L_{2}/r_h,\,r_+/r_h)$,\footnote{The results  qualitatively do not change even in case we do not normalize $L_{2}$ at all.} $(S,\,r_+/r_h)$ and $(S,\,L_{2}/T)$. In the three first plots the fraction $r_+/r_h$, align the results we want to compare in the horizontal axis. This happens because as we mention above the $r_+$ has almost linear dependence on $r_h$ for the values we are interested. Additionally, another reason for this choice is that the minimal surface has boundary always at $r_h$ and we would like to compare the difference in the extension of the world-sheet for varying $\xi$ with the other quantities.

From the $(L_{2}/T,\,r_+/r_h)$ results we study how deep in the bulk goes the world-sheet, for a soft gluon with particular energy. We would also like to see if there is still an upper bound for $L_{2}$ as happens in the case of zero chemical potential. We find that indeed it exists and we can calculate this upper bound with respect to $T$ (or $r_h$).

In order to examine the behavior of the corresponding minimal surface of our solutions we plot the results of $(S,r_+/r_h)$ and $(S,L_{2}/T)$. We could also divide the $L_2$ by $r_h$, or just plot the $L_{2}$, since the qualitative behavior of the results does not change. The first plot give us the amplitude in terms of $r_+/r_h$, and also is the criterium of which branch of values is the physically correct, ie. $r_+>r_{+Lmax}$, or $r_+<r_{+Lmax}$. The second plot, gives the minimal surface in terms of the energy of the soft gluon and allow us to compare them  at different chemical potentials. We remark here that we consider the regularization parameter in the integrations independent of any other parameter of the systems and we keep it always constant ie.  $\epsilon=10^{-5}$.

We intend to calculate the amplitudes by fixing the temperature to a large and small value and for each case by considering two different values $\xi_1$ and $\xi_2$.  For low temperature the values we choose for $R=1$ are:
\ben
\xi_1=\frac{\pi }{2 \sqrt{2}}\simeq 1.11072 \,,\qquad &\mbox{and}&\qquad
\xi_2=\frac{3 \sqrt{2} \pi }{5} \simeq 2.66573
\een
for  $T=\sqrt{2}/\pi\simeq 0.45016$.\footnote{For $\xi_1$ we can calculate  $r_{h_1}=1.5$ and $\m_{1}=0.5$, while for $\xi_2$ we get $r_{h_2} \simeq 1.85472$ and $\m_{2}=1.2$.}

We solve the system of equations \eq{ch1} and \eq{ch2} and plot the results $(L_{2}/T,\,r_+/r_h)$, $(L_{2}/r_h,\,r_+/r_h)$ in Figure 3. As we can see in this figure higher values of $\xi$ allow higher values of $L_{2,\,max}$, which are \footnote{We calculate the values of $L_{2,\,max}$ up to a constant $2 \sqrt{2}$, in order to be able to make direct contact with the numbers seen in the plots.}
\ben\nonumber
&&L_{2,\,max_1}\simeq 0.53747=1.19396~ T=0.35831 ~r_{h_1}\\&&
L_{2,\,max_2}\simeq 0.71773=1.59440~ T=0.38697~ r_{h_2}~.
\een
For this temperature it seems that the maxima, occur almost for the same values of the $r_+/r_h$. We also observe from the plots that for the same values of $L_{2}$, the world-sheet is less extended for higher chemical potentials than to lower ones.

Like the case of the finite temperature $\cN=4$ super Yang-Mills we should choose one branch of $r_+$ solutions corresponding to  $L_{2}$, and the answer to this will be given from the relations $S(r_+/r_h)$ and $S(L_{2})$. From these functions we can also extract the behavior of the amplitudes for different chemical potentials. Hence in Figure 4 we plot the action depending on $r_+$ and $L_{2}$.

We see that the natural branch to choose for the acceptable solutions is the one with $r_+< r_{+L,\,max}$. This is expected from
the fact that when we turn off the chemical potential we should get the case of the $\cN=4$ super Yang-Mills finite temperature we examined in the previous subsection.

From the Figure 4, we see that higher chemical potentials result slightly higher action values for fixed values of $r_+/r_h$, and this can be easily seen from the plots close to the maximum. The plot $S(L_{2})$ gives us information on the amplitude dependence of the chemical potential. Higher values of the chemical potential lead to lower values of the on-shell action.
\begin{figure}
\centerline{\includegraphics[width=70mm,height=40mm]{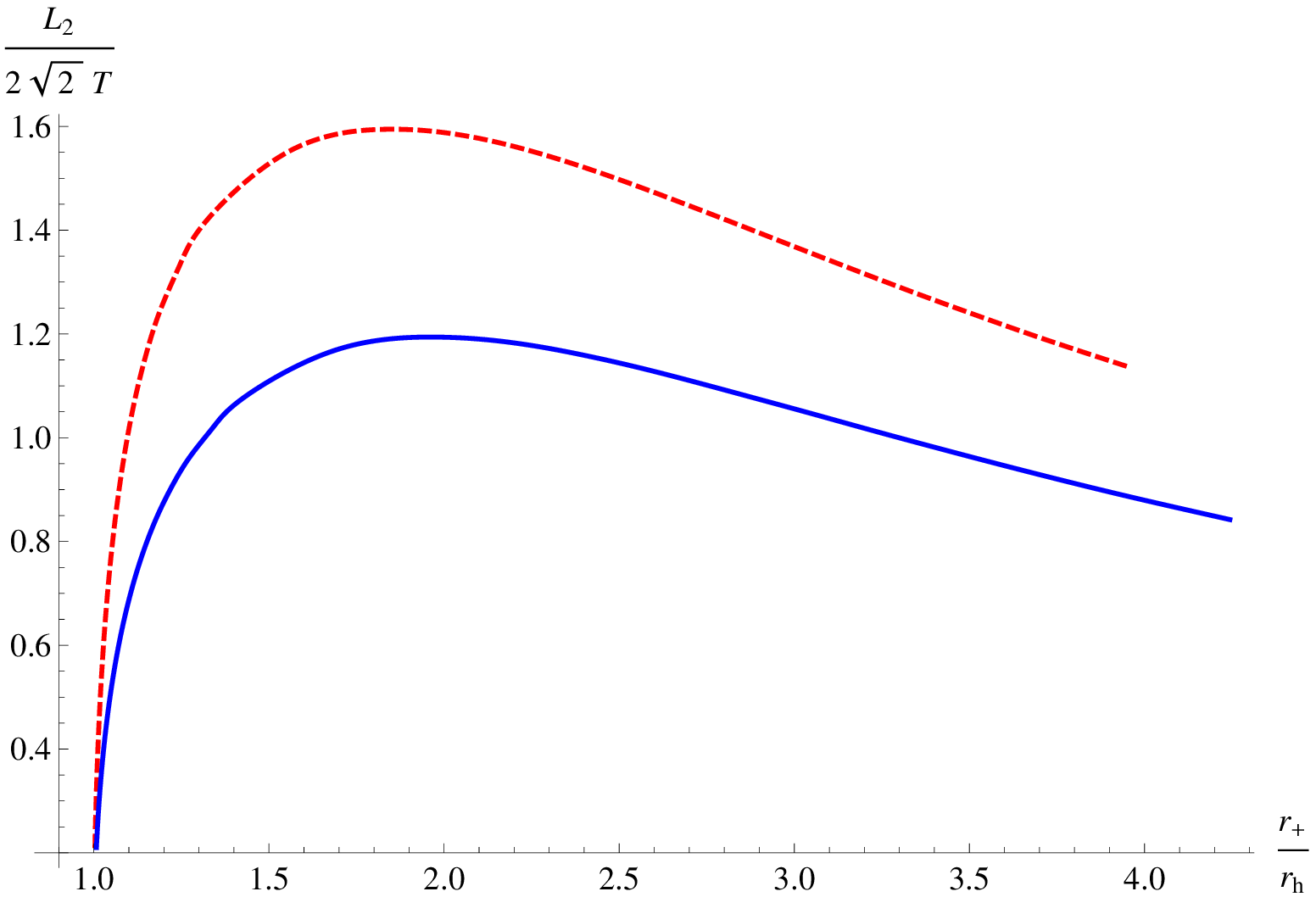}\qquad\includegraphics[width=70mm,height=40mm]{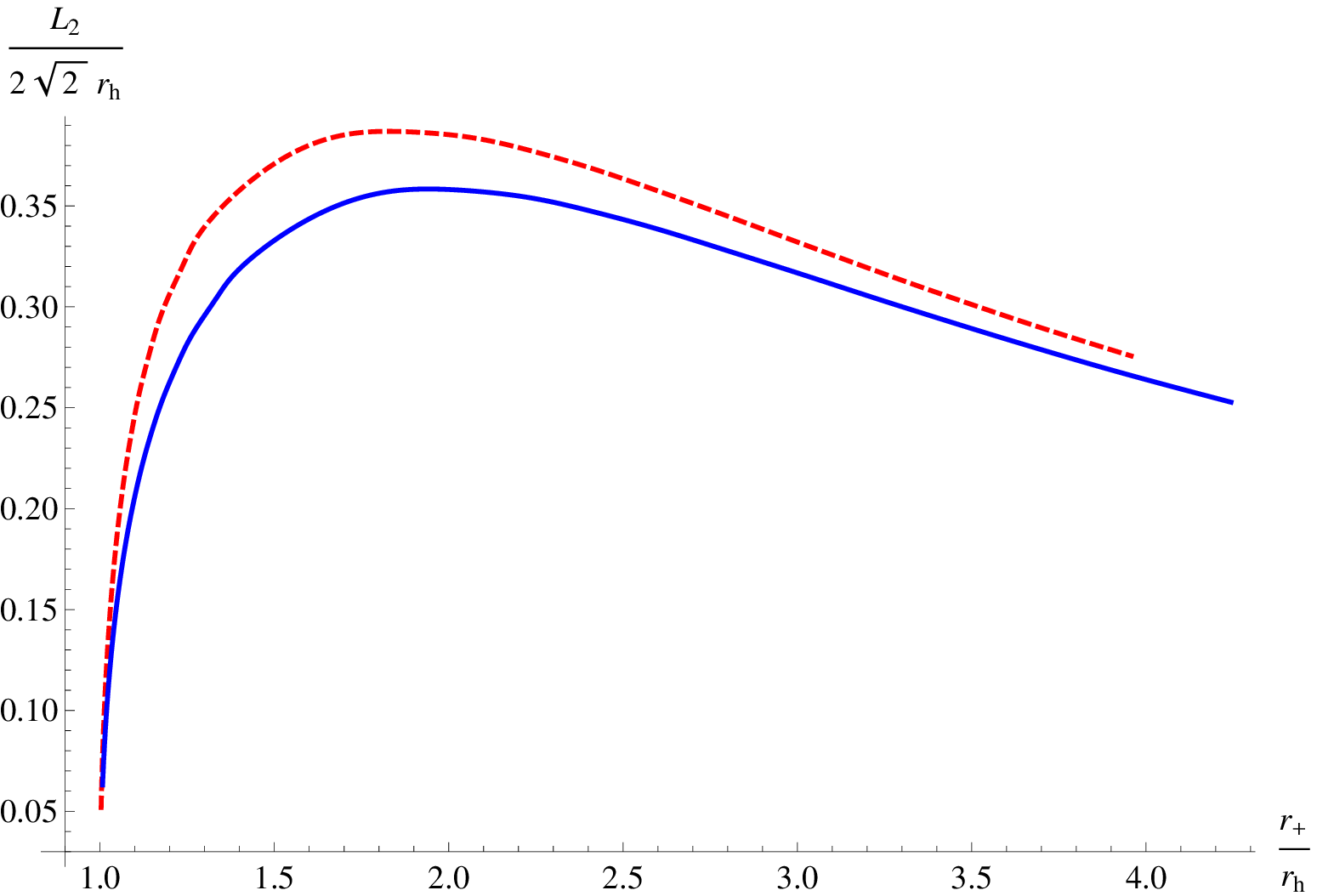}}
\caption{{\small The dependence of $L_{2}$ in terms of $r_+/r_h$ for $T=\sqrt{2}/\pi$. With red-dashed are the results for $\xi_2\simeq 2.66573$
and with blue the results for $\xi_1\simeq 1.11072$. Notice that for higher chemical potentials the $L_{2,max}$  takes higher values. On the left the $L_{2}$ is divided by $T$, while on the right by $r_h$. We observe that for the same values of $L_{2}$, the world-sheet is less extended in higher chemical potentials than in lower ones.}}
\centerline{\includegraphics[width=70mm,height=40mm]{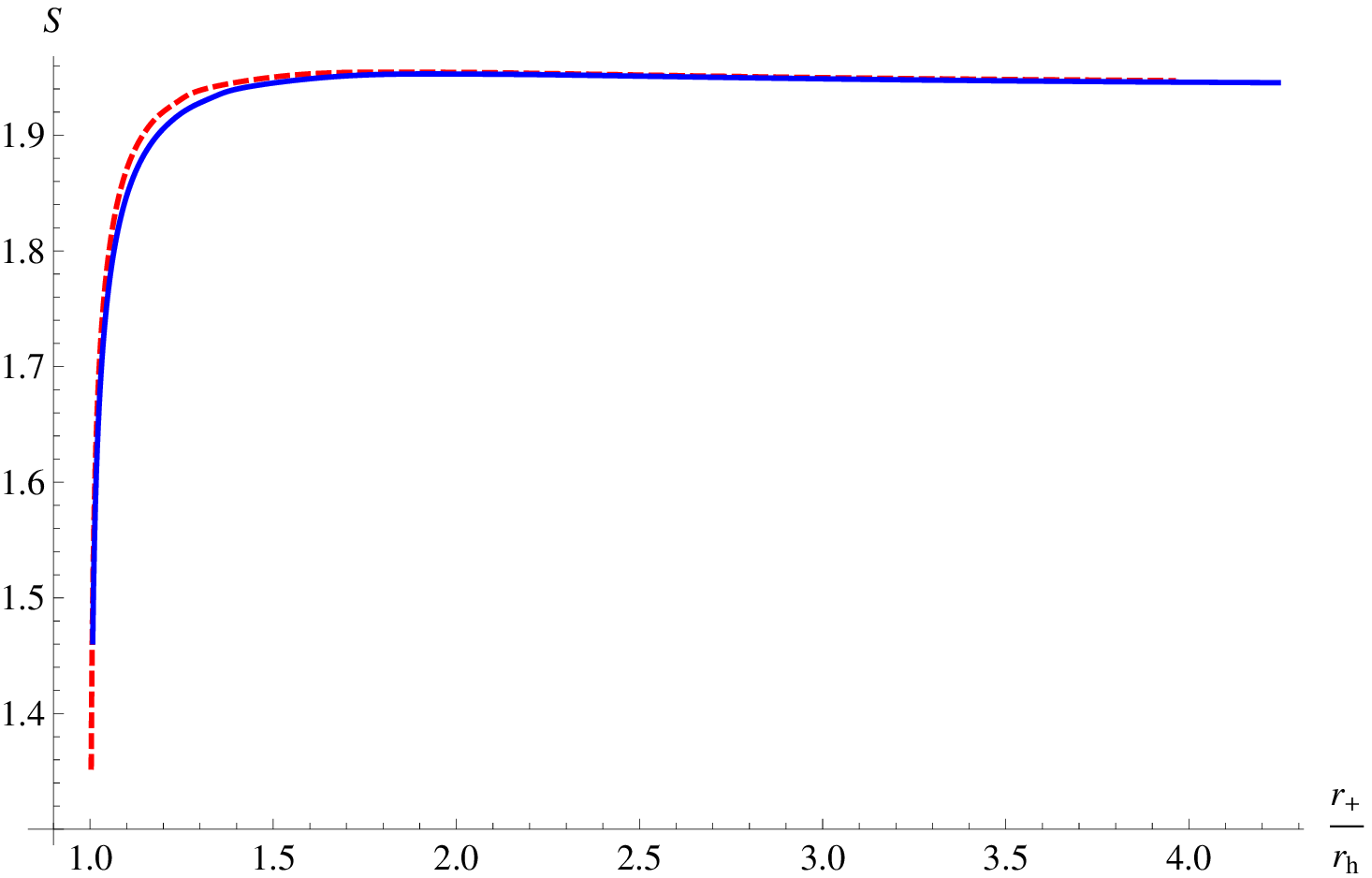}\qquad\includegraphics[width=70mm,height=40mm]{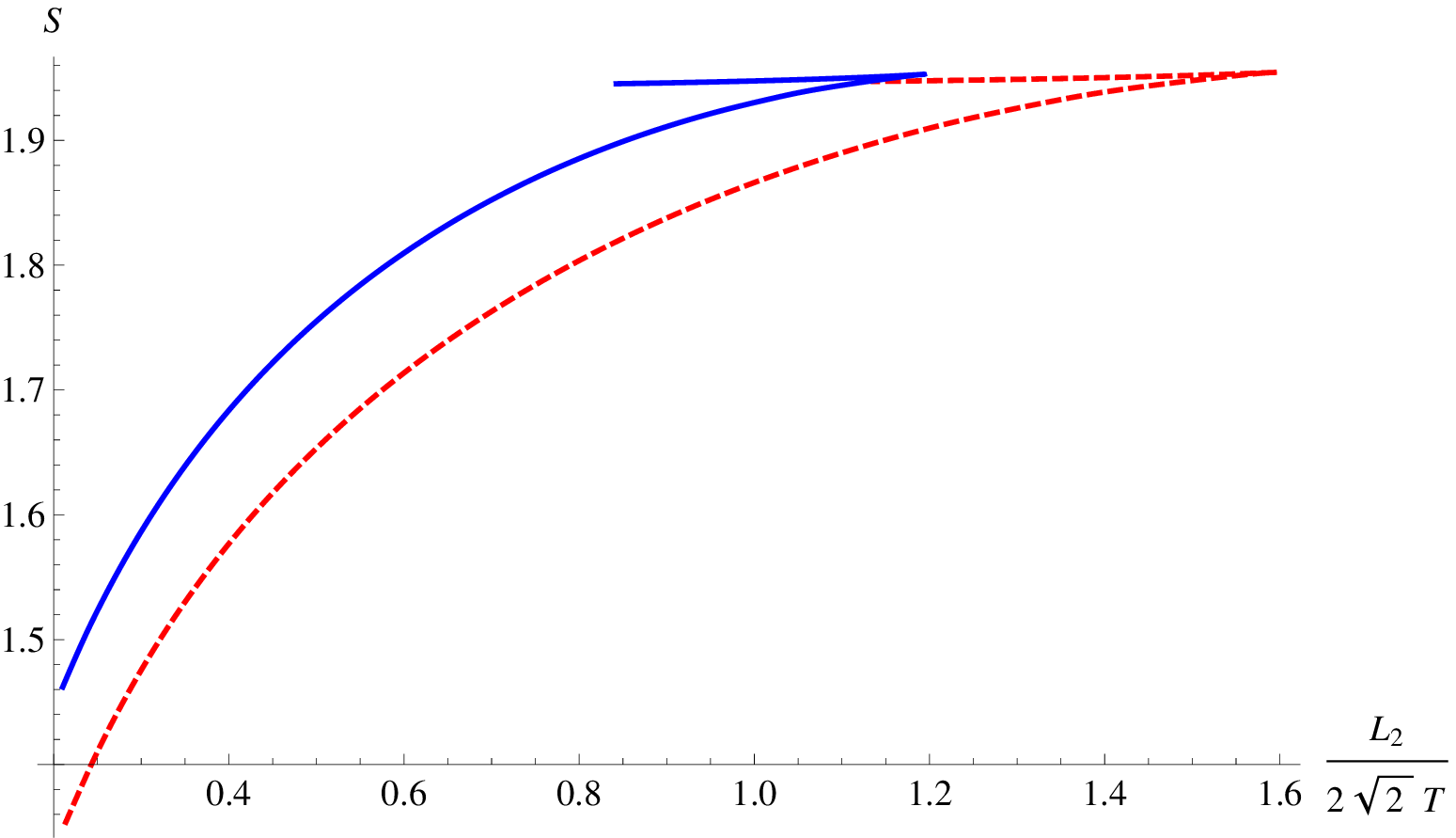}}
\caption{{\small On the left the dependence of the action as a function of $r_+/r_h$ for $T=\sqrt{2}/\pi$. The physical solution is the one that gives minimal action and corresponds to values of $r_+<r_{+L\,max}$. On the right, the dependence of the action on the $L_{2}/T$. For the same values of $L_{2}$ we get higher values of action for lower chemical potentials. For both of these plots the results corresponding to $\xi_2$ are plotted with red-dashed, while the ones corresponding  to $\xi_1$ are plotted with blue. }}
\end{figure}

\begin{figure}
\centerline{\includegraphics[width=70mm,height=40mm]{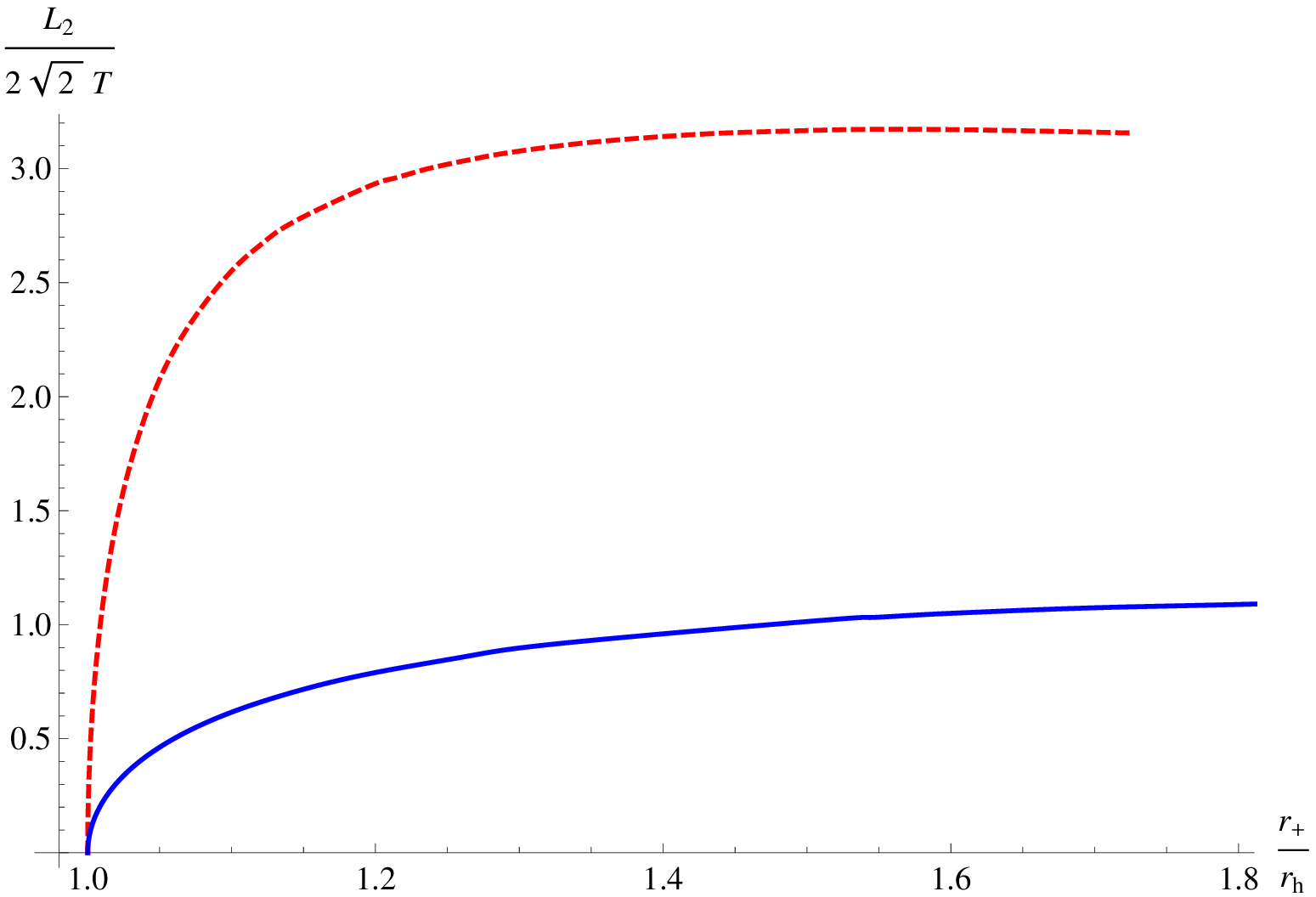}\qquad\includegraphics[width=70mm,height=40mm]{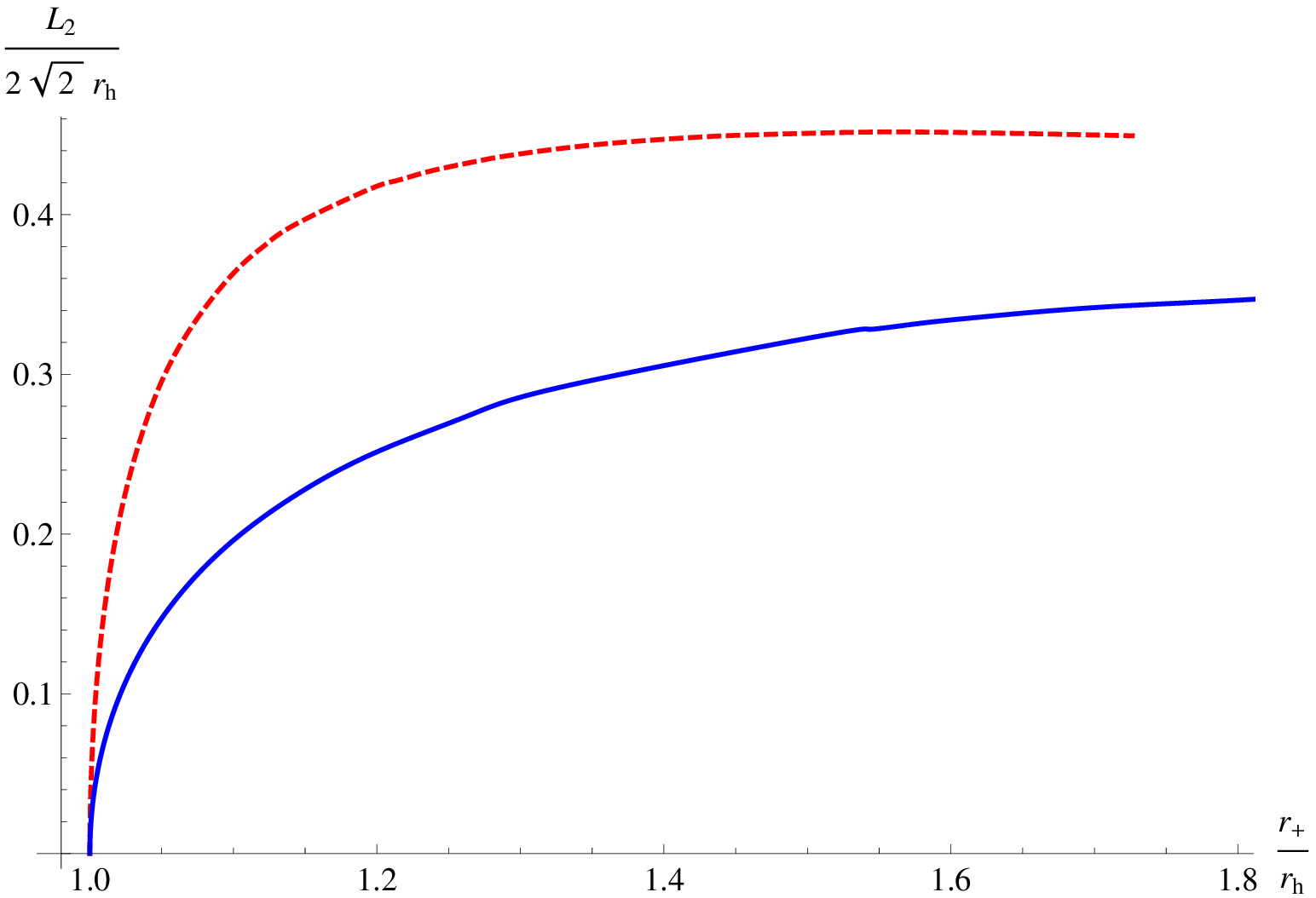}}
\caption{{\small The dependence of $L_{2}$ in terms of $r_+/r_h$ for $T=30/\pi$. With red-dashed are the results for $\xi_4=7.02481$ and with blue the results for $\xi_3=3.14334$. On the left the $L_{2}$ is divided by $T$, while on the right by $r_h$. We  observe that for the same values of $L_{2}$, the world-sheet is significantly less extended for higher chemical potentials than in lower ones, when there is an important difference between them.}}
\centerline{\includegraphics[width=70mm,height=40mm]{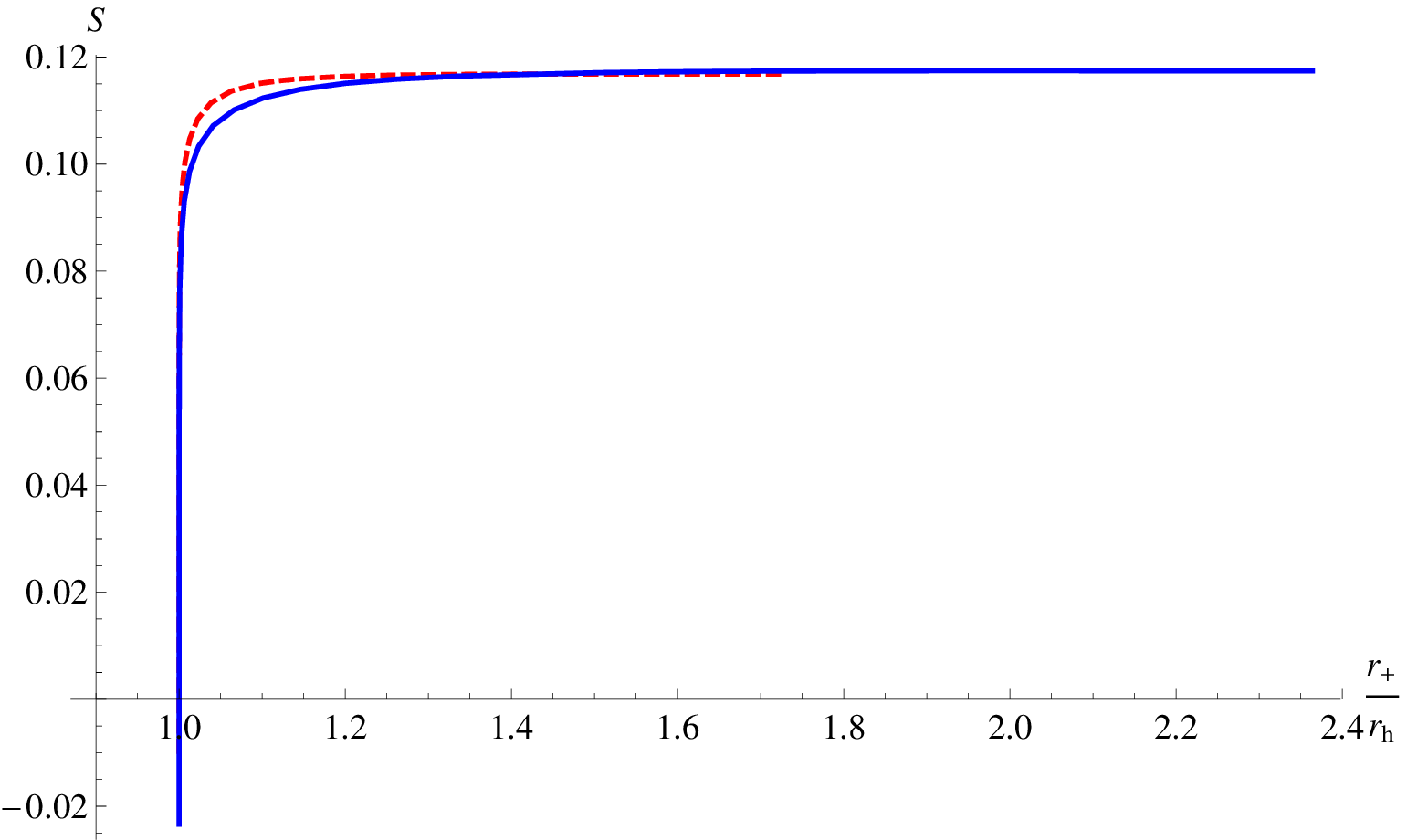}\qquad\includegraphics[width=70mm,height=40mm]{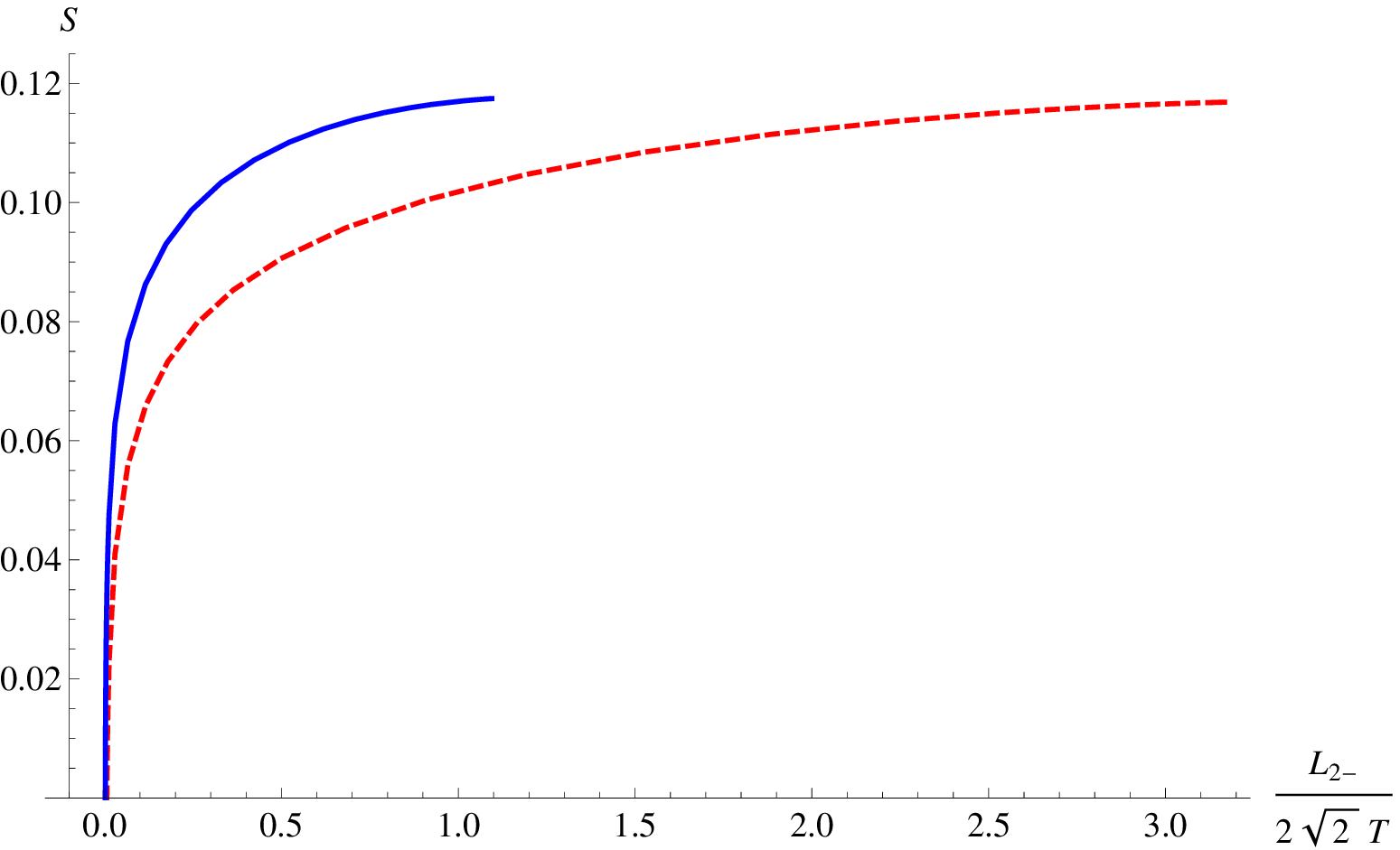}}
\caption{{\small On the left the dependence of the action as a function of $r_+/r_h$ for $T=30/\pi$. The physical solution is the one that gives minimal action and corresponds to values of $r_+<r_{+L\,max}$. On the right, the dependence of the action on the $L_{2}$. The cusp we observed in the previous cases is still there, but is not easily seen on the plot. More points however in this region would make it appear clearly. Comparing to the Figure 4, we see that although we consider $\xi_4-\xi_3$ significantly bigger than $\xi_2-\xi_1$ the difference in the scattering amplitudes for the chemical potential does not change significantly. For both of the above plots the results corresponding to $\xi_4$ are plotted with red-dashed, while the $\xi_3$ results are plotted with blue.}}
\end{figure}
We would also like to examine the amplitude dependence on the chemical potential for a higher value of temperature choosing, $T=30/\pi\simeq 9.5493$. In that case we can choose values for $\xi_i$  which differ significantly, so
\ben
\xi_3=\frac{\sqrt{901}\pi}{30}\simeq 3.14334 \qquad &\mbox{and}&\qquad \xi_4=\sqrt{5}\pi \simeq 7.02481~
\een
and as a result the chemical potentials for this choice differ significantly\footnote{For $\xi_3$, the corresponding values for the horizon and the chemical potential are $r_{h_3}=\sqrt{901}$ and $\m_{3}=1,$ while for $\xi_4$  the $r_{h_4}=\sqrt{4500}$ and the $\m_{4}=60~$.}.

We again solve the system of equations and plot the results. Looking at the Figure 5, and compare the results  with the ones in Figure 3, we see that higher differences on the chemical potential lead to increasing differences in the $L_{2,max}$ value. We also see that the corresponding $r_{+Lmax}/r_h$ value, seems to be slightly increasing as the chemical potential reduces. Hence for particular values of $L_{2}$ the corresponding string world-sheet is extended more for lower chemical potentials and as the difference between the chemical potentials increases, the difference of how much extended the corresponding world-sheets are, increases.

From Figure 6, we conclude that for the same $L_{2}$, the differences of the on-shell action for increasing differences between the  chemical potentials are increasing. Hence what matters is the difference in the parameters $\xi_i$ and not the fraction of them, since $\xi_2/\xi_1>\xi_4/\xi_3$ but $\xi_2-\xi_1<\xi_4-\xi_3$.

Before we conclude this section, it would be good idea to fix the $L_2$ and the temperature, and modify the dimensionless parameter $\xi$.
Doing that we would be able to observe how the argument of the complex amplitude is modified with $\xi$ for a fixed soft gluon momentum and temperature.  The numerical analysis is more difficult here but still doable.

So we choose two temperatures, $T_a=0.5$ for $L_{2a}\simeq 0.128$ and $T_b=10$ for $L_{2b}\simeq 0.112${}\footnote{We have chosen the values $L_{2a}$, and $L_{2b}$ to be close, but also in a way that is computationally convenient to find the corresponding results.} and plot the action S in Figure 7. We observe, that for the two different temperatures and fixed $L_2$ the action $S$ does not vary significantly, while varying $\xi$ from its minimum to its maximum value. We also see that the on-shell action is everywhere finite as expected since even for zero chemical potential it has non zero value. We also observe that in the two plots the function $S(\xi)$ seems to have similar form. However, for increasing temperature the action takes lower values for fixed chemical potential.
\begin{figure}
\centerline{\includegraphics[width=70mm,height=40mm]{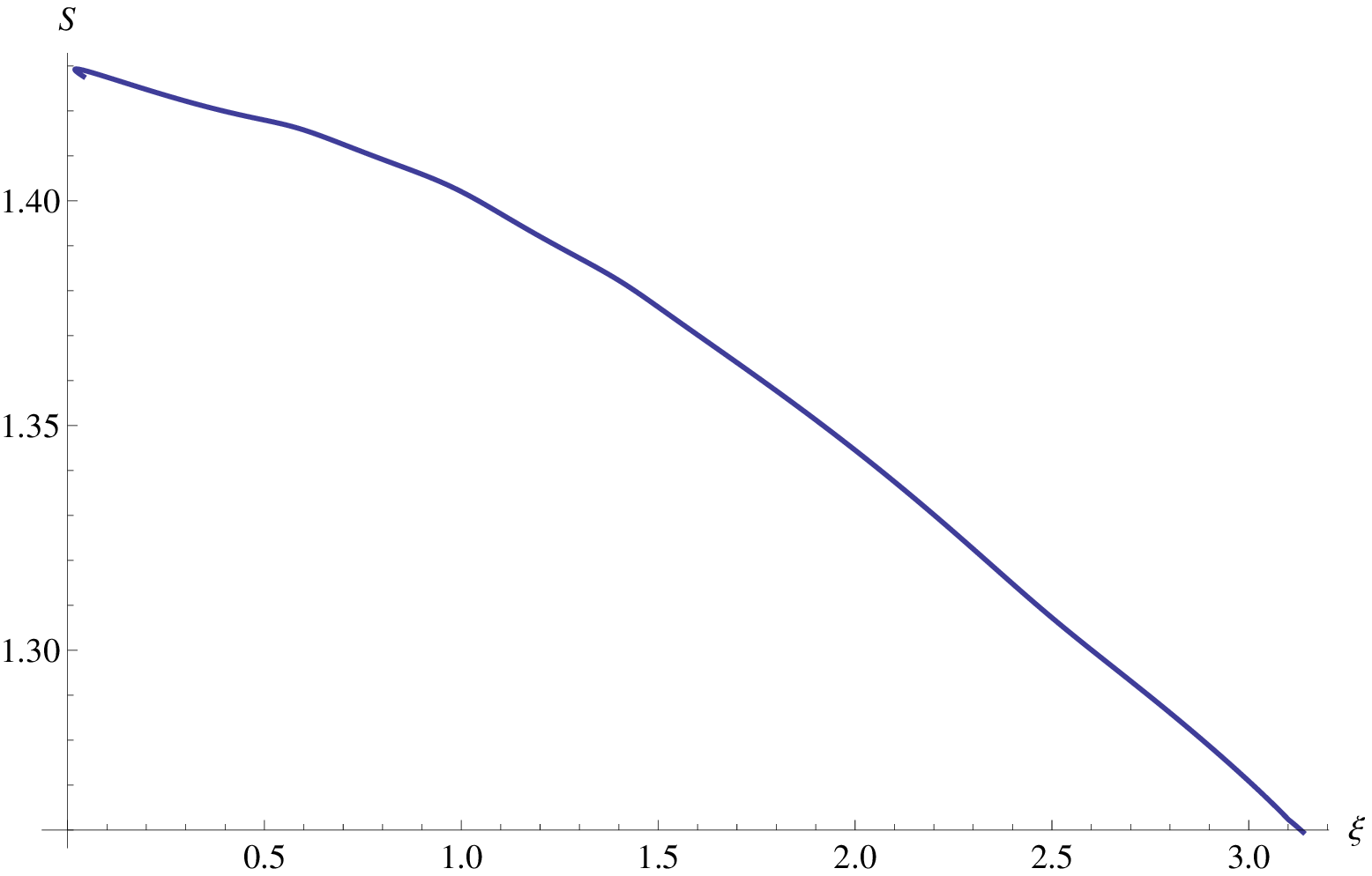}\qquad\includegraphics[width=70mm,height=40mm]{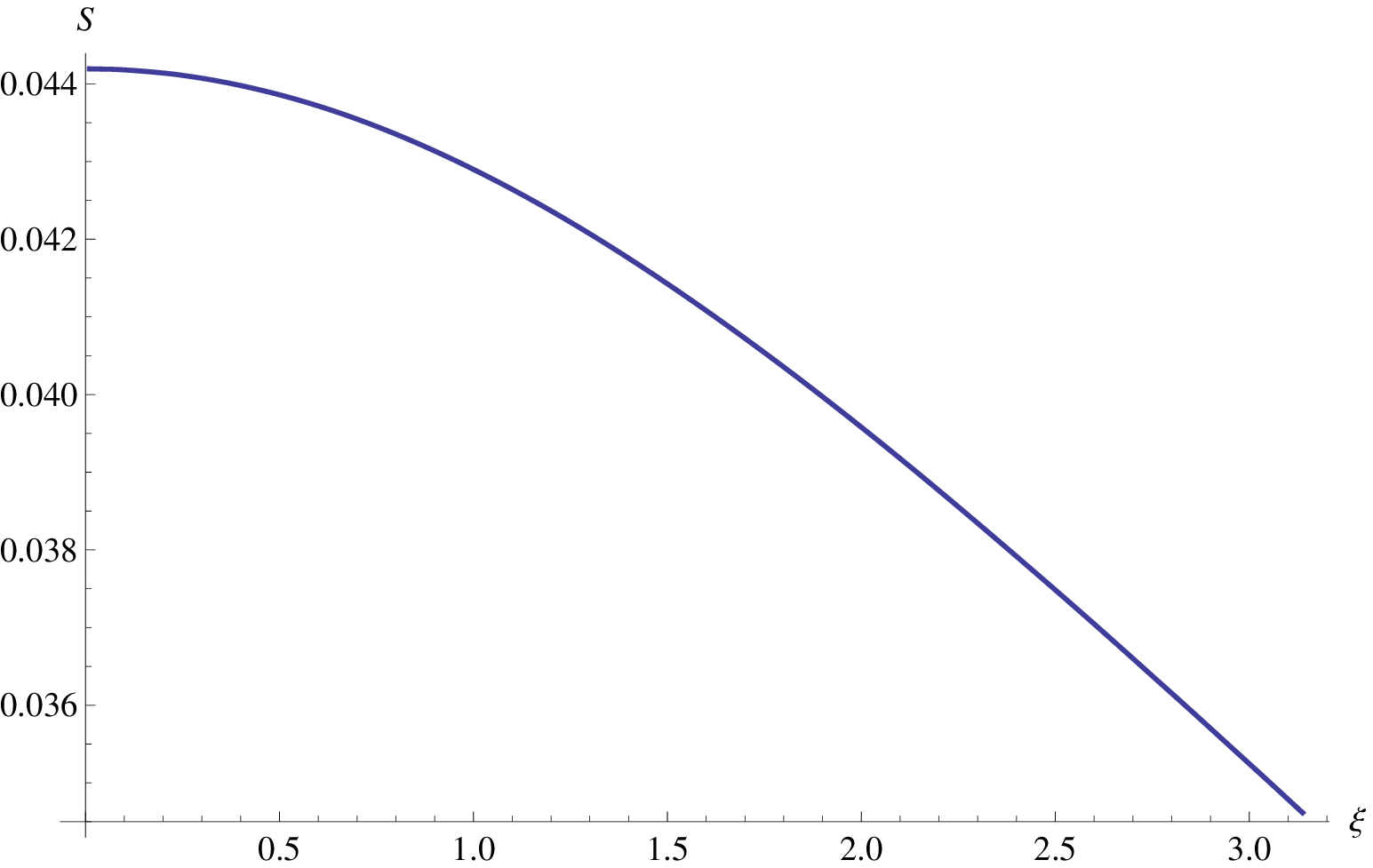}}
\caption{{\small The on-shell action, depending on $\xi$ for a fixed length $L_2$ and temperature. For the left plot we choose $T_a=0.5$ for $L_{2a}\simeq 0.128$ and for the right one $T_b=10$ for $L_{2b}\simeq 0.112$}}
\end{figure}

We briefly summarize some of our findings in this section. We examined the dependence of the gluon scattering amplitude on the chemical potential, keeping fixed the temperature and we find that higher chemical potentials lead  to lower on-shell actions for the Wilson loop we consider. The increase that happens depend on the absolute difference between the parameters $\xi_i$ and not on their fractions. This gluon scattering amplitude coming from the calculation of the AdS/CFT is a phase since the Wilson loop has Lorentzian signature.


\subsection{Gluon scattering amplitude in finite temperature non-commutative gauge theories}

In this subsection we examine the gluon scattering amplitudes in the non-commutative finite temperature gauge theories \footnote{The scattering amplitude in the zero temperature non-commutative gauge theory have been examined in \cite{Severnc}.}. For this setup we need to do a T-duality to the four coordinates of \eq{metricnc} to get the
background \eq{nctmetric}.
The ansatz for the string world-sheet is
\be
y_0=\tau +z(\s)~,\qquad y_2=\t~,\qquad y_3=\sigma~, \qquad r=r(\s)
\ee
and we use the boundary conditions of \eq{ybc}. This choice of ansatz
allow us to have only one time. Notice that in the ansatz we have renamed the function of $\s$ in $y_0$ to avoid confusion with the function $f$ appearing in this background.
The action becomes
\ben\nonumber
&&S=\frac{R^2}{2 \pi\a'}\int dy_1 dy_2\Big[\frac{1}{r^2}\Big(\frac{(\partial_0 y_0)^2(\partial_1 y_0)^2}{A_0^2}-\\ \label{actionnc}
&&\qquad\qquad\qquad\qquad\qquad\left(1-\frac{(\partial_0 y_0)^2}{A_0}\right)\left(1-\frac{(\partial_1 y_0)^2}{A_0}+\frac{r'^2}{f}\right)\Big)\Big]^\frac{1}{2}-a^2~,
\een
where
\be
A_0=h(f+a^4 r^4)~.
\ee
The equations of motion for $z$ give
\be\label{znc}
z'=A_0\sqrt{D} r^2 c_1~,
\ee
where
\be
D:=\frac{1}{r^2}\left(\frac{(\partial_0 y_0)^2(\partial_1 y_0)^2}{A_0^2}-\left(1-\frac{(\partial_0 y_0)^2}{A_0}\right)\left(1-\frac{(\partial_1 y_0)^2}{A_0}+\frac{r'^2}{f}\right)\right).
\ee
Going to the Hamiltonian formalism, we calculate the conjugate momenta
\be
p_z=\frac{z'}{r^2 A_0 \sqrt{D}}~,\qquad p_r= \frac{r'}{r^2\sqrt{D} f}(1-\frac{1}{A_0})
\ee
and the Hamiltonian becomes
\be\label{hnc}
\cH=\frac{1}{r^2\sqrt{D} }(1-\frac{1}{A_0})+a^2~,
\ee
where the first term is negative for any value of $r$.
Considering $y_2$ as a time we can set the Hamiltonian constant and equal to $1/\b_0$.
From the form of the Hamiltonian \eq{hnc} we get a first constraint for the non commutative parameter and the constant
$\b_0$, which is $a^2>\b_0^{-1}$, since $0<A_0<1$.
We can solve \eq{hnc} for $\sqrt{D}$ and get
\be\label{Dnc}
\sqrt{D}=\left(\frac{1}{\b_0}-a^2\right)^{-1}\left(1-\frac{1}{A_0}\right)\frac{1}{r^2}
\ee
and substituting to \eq{znc} we obtain a relatively simple equation for
$z'$
\be\label{zznc}
z'=\frac{c_1}{\b}(A_0-1)~,
\ee
where we set
\be\label{betaca}
\b:=1/\b_0- a^2~,
\ee
Using \eq{Dnc} and \eq{zznc} we derive the turning point equation for the world-sheet
\ben\nonumber
r'{}^2&=&\left(\frac{1}{\b^2}\left(1-\frac{1}{A_0}\right)\left(c_1^2 A_0-\frac{R^4}{r^4}\right)-1\right)f\\\label{uprimenc}
&=&\left(1-\frac{r_h^4}{r^4}\right) \left(\frac{c^2 r_h^4 \left(\left(1+a^4 r^4\right) \left(R^4-c_1^2 u^4\right)+c_1^2 r_h^4\right)}{\left(a^2 c-1\right)^2 r^4 \left(1+a^4 r^4\right) \left(r^4+a^4 r^8-r_h^4\right)}-1\right)~
\een
and is not difficult to see that for $a=0$ this equation reduces to \eq{rprime} as it should be. The equation \eq{uprimenc} can be reduced to a fourth order algebraic equation\footnote{We ignore the multiplicative factor $f$ in the equation since it contributes four solutions $r=r_h$, which are not acceptable in our case.} and can be solved analytically. The turning point solution can be found and is a lengthy function which depends on $\b,\,a,\,u_h,\,R$ and $c_1$ and we denote the acceptable solution as $r_{+}$.

Hence we end up with a system of two equations with two unknown constants $\b,\,c_1$, \footnote{Equivalently we could consider as unknown parameters the integration constants $\b_0$ and $c_1$.} which can  be solved arithmetically and give $L_{2}$ as a function of $r_{+}/r_h$.
These two integrals are obtained from \eq{zznc} and \eq{uprimenc} respectively
\ben
 \frac{L_{2}}{2 \sqrt{2}}&=&\frac{c_1}{\b}\int_{r_h}^{r_{+}} dr\frac{A_0-1}{r'}~,\label{l2nc2}\\\label{l2nc}
 \frac{L_{2}}{2 \sqrt{2}}&=&\int_{r_h}^{r_{+}} \frac{dr}{r'}~,
\een
where $r'$ is given from equation \eq{uprimenc}.
Unfortunately the integrals cannot be calculated analytically.
So one can think to expand them for small values of $a$, keep the dominant terms and then try to solve the system analytically. But expansion up to order $a^2$ makes the whole problem equivalent to the commutative one, just by redefining the constants as
\ben
\hat{\b}_0^2:=(1-2 a^2 \b_0) \b_0^2\qquad \mbox{and}\qquad \hat{c}_1^2=c_1^2~,
\een
where $\hat{\b}_0$ is the inverse of the Hamiltonian and $\hat{c}_1$ is the constant introduced from the integration of the function $z(\s)$.
By including higher order terms the integrals can not be solved analytically.

So we need to do the analysis arithmetically. For convenience we set
\be
R=r_h=1
\ee
and we will give several different values to the non-commutative parameter $a$, in order to examine the non-commutative dependence of the gluon scattering amplitudes. We choose a value that makes the commutative effects almost negligible, $a=0.01$, one intermediate value $a=0.3$ and one bigger $a=1$ where we expect the non-commutative effects to modify significantly the relevant commutative results.

For each value of the non-commutative parameter, we solve the system of  equations \eq{l2nc2}, \eq{l2nc} and find the pair of parameters $\b,\,c_1$ for the corresponding values of $L_{2}$.
Then we plot $L_{2}$ as a function of $r_+/r_h$ for different values of the non-commutative parameter and the results appear in the Figure 8.
\begin{figure}
\centerline{\includegraphics[width=80mm]{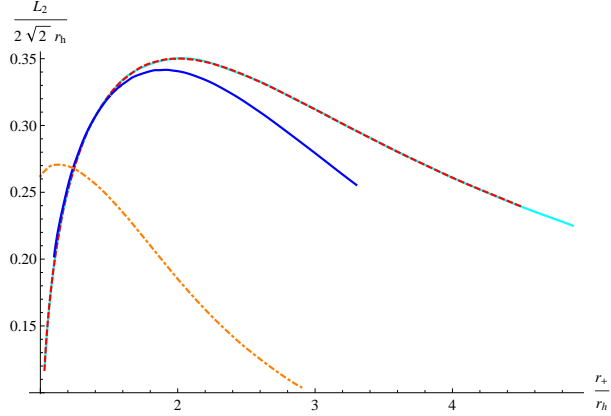}}
\caption{{\small  $L_{2}/2\sqrt{2} r_h$ as a function of $r_+/r_h$ for different values of the non commutative parameter. The cyan-continuous color corresponds to the commutative case, and as we see the curve almost coincides  with the non commutative case for the very small commutative parameter $a=0.01$, which is plotted with red-dashed. The commutative effects in the gluon scattering amplitude begin to become important for $a=0.3$ and which is plotted with solid blue color. The values of $a=1$ are plotted with the orange-dotdashed color and although the qualitative form of the figure remains the same with the commutative case the quantitative results change significantly.}}
\end{figure}
For $a=0.01$ we observe that the non-commutative effects are negligible. For bigger values of $a$ the results become very interesting. For $a=0.3$
we see the maximum value of $L_{2}$ reduces with respect to the commutative case and occurs for smaller value of $r_+/r_h$.
The effects of non-commutativity become very important for $a=1$. The maximum value of $L_{2}$ reduces approximately to $77$ percent of
the commutative value and this maximum occurs at the half of the relevant commutative value for the turning point. We also observe that the increasing branch of the curve has starting point for $L_{2}/2\sqrt{2}$ bigger than $\simeq 0.26$. That is interesting and means that there exist single solutions $r_+$ for $L_{2}/2\sqrt{2}<0.26$.

Of course the question that one has to ask again is which branch is the physical one, and whether or not there is a 'jump' in the turning point solutions from one branch to another. To answer this question we need to derive the regularized action
\be
S=\frac{L R^2}{\sqrt{2}\pi\a'}\left(\int_{r_h}^{r_+}\frac{r_h^4}{r^2}\frac{1}{\sqrt{\b^2  \left(r^4+a^4 r^8-r_h^4\right)} r'}dr-a^2 \frac{L_2}{\sqrt{2}}\right)~,
\ee
where $r'$ is given by \eq{uprimenc}.
We calculate the action for the three values of
the non commutative parameter $a$ and using the triplet values  $(\b,\,c_1,\,L_{2})$ from the solution of the system \eq{l2nc2} and \eq{l2nc} we find that the physical branch is the left one, at least for non-commutative parameter values $a=0.01,\,0.3$ where two solutions of $r_+$ exist for each $L_2$. The results are plotted in Figure 9. However the case of $a=1$ is more complicated. We see that for values of $L_{2}/2\sqrt{2}$ smaller than $\simeq 0.26$ the only solutions that exist are in the right decreasing branch of the Figure 8.

Thinking like the previous cases, we should consider as the acceptable solutions the ones that have $r_+<r_{+Lmax}$. However if we set as additional criterium for choosing the acceptable solutions the minimum of the action, then for Wilson loops with edges approximately $0.26<L_{2}/2\sqrt{2}<0.271$ the physical branch is the left one for $r_+<r_{+Lmax}$. For approximately $L_{2}/2\sqrt{2}<0.26$ the acceptable solutions are for $r_+>r_{+Lmax}$ which happens for $r_{+}\simeq 1.33 r_h$.

Geometrically this means that by starting with a short edge $L_{2}$ and increasing it, we get a world sheet that has a turning point that gets closer to the boundary in a continuous way. However, when we reach $L_{2}/2\sqrt{2}\simeq 0.26$ the turning point from $r_+\simeq 1.33 r_h$ goes to   $r_+\simeq 1.01 r_h$ and as we continue increasing $L_{2}$ the turning point now is going deeper to the bulk until the edge gets its maximum value $L_{2max}/2\sqrt{2}\simeq 0.271$ which corresponds to $r_{+Lmax}\simeq 1.15 r_h$. However, notice that in the limit $L_2\rightarrow 0$ the solutions for $r_+>r_{+Lmax}$ can not be acceptable.

The on-shell action in terms of $r_{+}/r_{h}$ is increasing as we decrease the noncommutative parameter. Around the 'jump' of the solutions we mentioned,  the action varies steeper for the values  $r<r_{+Lmax}$ occurs. This can be also seen from the Figure 9.

\begin{figure}[t]
\centerline{\includegraphics[width=75mm]{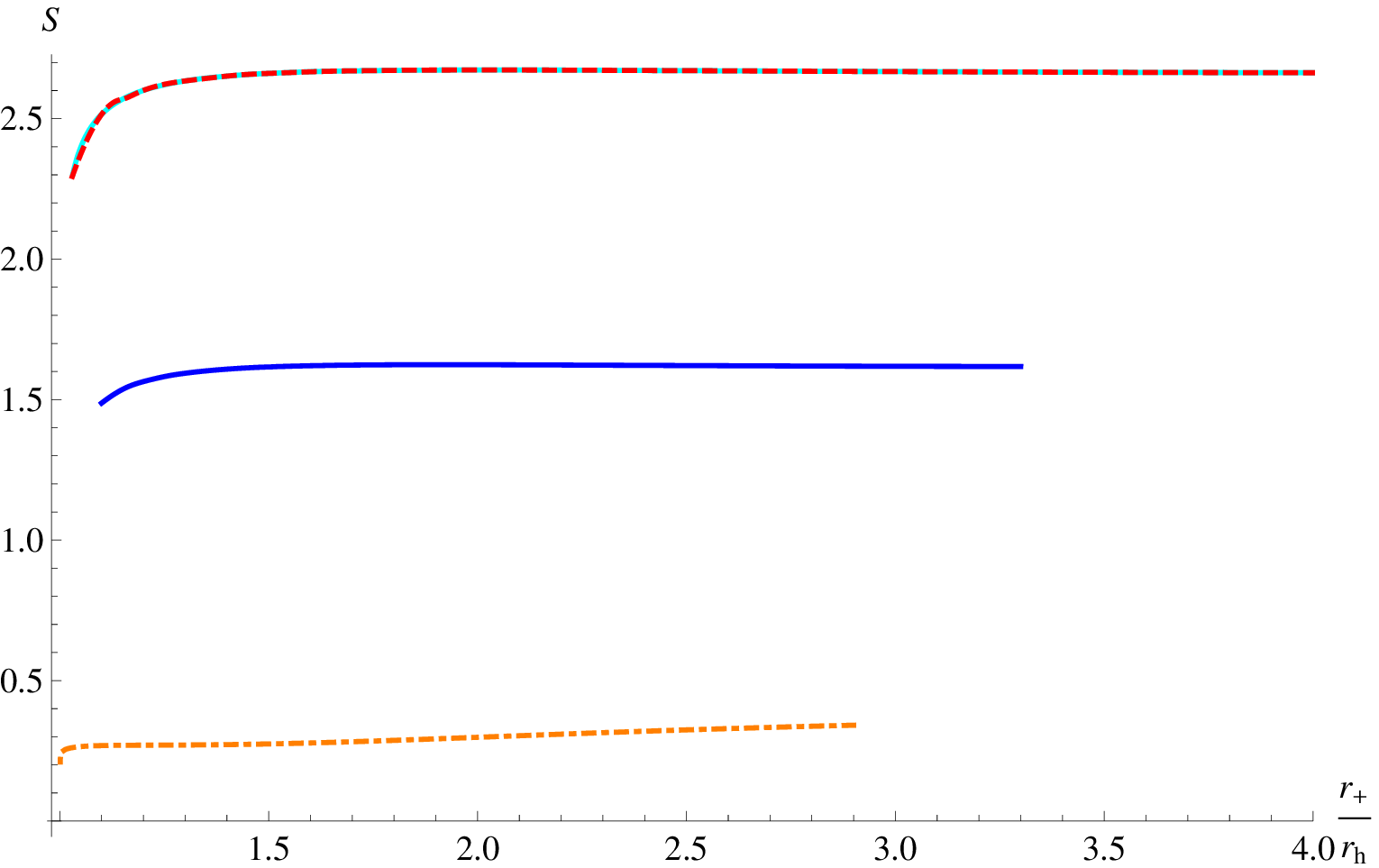}\quad\includegraphics[width=75mm]{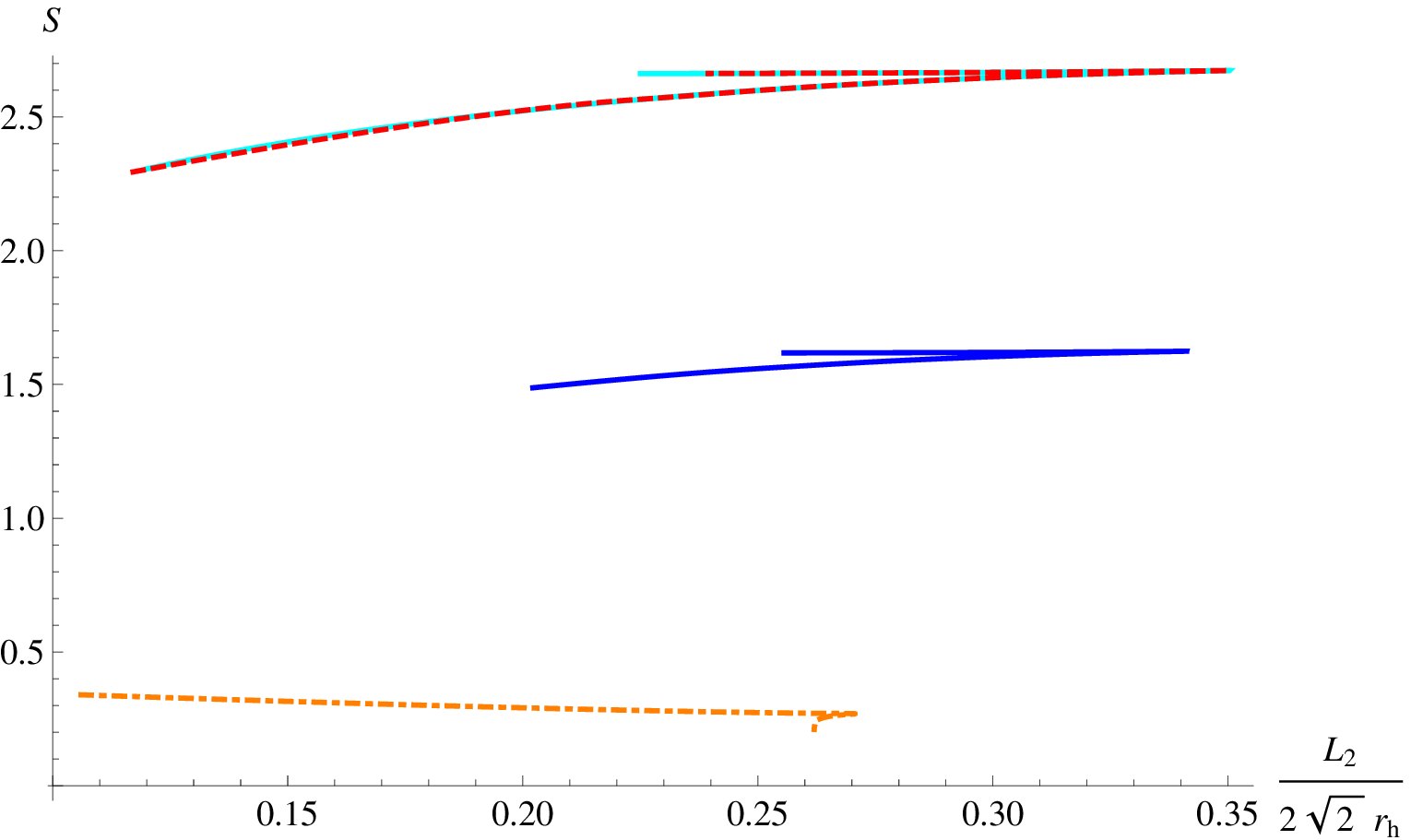}}
\caption{{\small On the left, the action for different values of the noncommutative parameter. The colors are as in the previous figure ie. ($a=0$, continuous cyan), ($a=0.01$, dashed red), ($a=0.3$, solid blue) and ($a=1$, dotdashed orange) colors. The maximum of the curve moves to the left as the noncommutative parameter is increasing. This can be seen from the Figure 8 too. The on-shell action is decreasing as the non commutative parameter is increasing. On the right,  $S$ as a function of $L_{2}$ for different values of the non commutative parameter. We see that the physical branch is the one with the smallest energy corresponding to the left one of Figure 8. However we note also here that for $a=1$ there exist unique solutions for small values of $L_{2}$ that belong to the right branch of Figure 8, i.e $r+>r_{+Lmax}$.}}
\end{figure}
In this section, we have examined the minimal surface ending at a particular light-like Wilson loop which is located at the boundary of a T-dual black hole in a background with B-fields, corresponding to a non-commutative gauge theory at finite temperature.
We find that by increasing the non-commutative parameter, the on-shell action is decreasing. Moreover our solutions are constrained, in a sense that the short edge of the Wilson loop is constrained by an upper bound depending on the temperature and the non-commutative parameter. For these solutions we observe a property that did not exist in the other theories we examined, and is a non-continuous 'jump' of the turning point of the world-sheet as we  vary the short edge of the loop.  As we already mentioned we believe that the considered Wilson loop corresponds to a gluon scattering amplitude of a low energy gluon off a high energetic in the non-commutative gauge theory at finite temperature.

\section{Discussion}

In this paper we examined the gluon scattering amplitude in finite temperature gauge/gravity dualities. We define the amplitude as the
minimal surface ending at the Wilson loop on the horizon of the T-dual back hole. Because of the computational difficulties we did not manage
to calculate the most general 4-gluon scattering amplitude. However, we have constructed the minimal surface of
a light-like Wilson loop with one edge
much more bigger than the other, and we claim, as in \cite{Nastasegs}, that this corresponds to an amplitude describing the forward scattering of a
low energy gluon off a high energy one.

We chose consistently the boundary conditions and the ansatz for the world-sheet, and found the characteristics of these solutions
and their corresponding minimal surfaces in the three theories we considered. It turns out that in all three cases our solutions are constrained.
Namely, the short edge of the Wilson loop is constrained by an upper bound which depends on the temperature, the chemical potential
and the commutative parameter. We find that for increasing chemical potential this upper bound increases, while increasing non-commutative parameter
leads to a decrease of the upper bound.

In general we see that for each value of the short Wilson loop edge $L_2$, there are two solution for the turning point equations.
It turns out that the acceptable one for the gluon scattering amplitude interpretation, is the one where the string world-sheet stays closer to
its boundary. Hence the maximum allowed penetration length $r_{+Lmax}$ of the world-sheet is where the $L_2$ becomes maximum. An increase
of the chemical potential leads to a slight increase of this quantity, while an increase in the non-commutative parameter leads
to a decrease of $r_{+Lmax}$.

Another characteristic of our solutions is that for the same Wilson loops, ie. same $L_{2}$, the minimal surfaces are more extended
in the T-dual bulk for lower chemical potential. This is one of the reasons that the corresponding minimal surface, for the same values
of the $L_{2}$ and increasing chemical potentials has decreasing area or equivalently decreasing on-shell action. The on-shell action
in the non-commutative theory has similar behavior, where the increase of the non-commutative parameter leads to the decrease of the on-shell action.

At the limit of zero chemical potential, and of zero non-commutativity, our results reduce smoothly to the finite temperature results.
However, the limit of zero temperature for our solutions does not exist since at this limit the upper bound of $L_{2}$ becomes zero
and the whole setup we considered becomes inappropriate.

For the minimal surfaces of all the four point amplitudes we calculated, we have used an appropriate cut off which corresponds
in the field theory in the IR cut-off regularization. This is consistent, since it is equivalent of using the Legendre transform
of the  action in a formal treatment, since the background satisfy the conditions of \cite{giataganas08}. Although the amplitude
for our kinematic configuration turns out to be a pure phase,
it is interesting to notice that the minimal surfaces we found have the properties that one would expect for a real amplitude
of the form  $\exp(-S)$ in finite temperature gauge theories. Currently we do not know if this is just a coincidence.

It would be very interesting to consider the most general planar four point gluon scattering amplitude using the corresponding
general polygonic sequence where we expect that the exponent in the amplitude should have a real part too.  Moreover by achieving that,
we would be able to  directly calculate the viscosity coefficient at finite temperature.

\textbf{Acknowledgements:} We would like to thank Katsushi Ito, Horatiu Nastase and Koh Iwasaki for useful correspondence.
DG would also like to thank Michael C. Abbott, Chong-Sun Chu, Robert de Mello Koch and Kevin Goldstein for useful discussions or correspondence.
The research of DG is supported by a SARChI postdoctoral fellowship. Part of this work was done while GG was at Queen Mary University.

\end{document}